\documentclass{emulateapj}

\begin{document}

\shortauthors{Gordon et al.}
\shorttitle{MIPS Reduction Algorithms}

\slugcomment{PASP, in press (May 2005)}

\title{Reduction Algorithms for the Multiband Imaging Photometer for
Spitzer} 

\author{Karl D.\ Gordon\altaffilmark{1}, George H.\ Rieke, Charles W.\ Engelbracht,
  James Muzerolle, John A.\ Stansberry, Karl A.\ Misselt, 
  Jane E.\ Morrison, James Cadien, Erick T.\
  Young, Herv\'{e} Dole, Douglas M.\ Kelly, 
  Almudena Alonso-Herrero, Eiichi Egami, Kate Y.\ L.\ Su, 
  Casey Papovich, Paul S.\ Smith, Dean
  C.\ Hines, Marcia J.\ Rieke, Myra Blaylock, 
  Pablo P\'erez-Gonz\'alez, Emeric LeFloc'h, Joannah L.\ Hinz} 
\affil{Steward Observatory, University of Arizona, Tucson, AZ 85721}
\and
\author{William B.\ Latter, Ted Hesselroth, David T.\ Frayer, 
  Alberto Noriega-Crespo, Frank J.\ Masci, Deborah L.\ Padgett}
\affil{Spitzer Science Center, Pasadena, CA 91125}
\and
\author{Matthew P.\ Smylie and Nancy M.\ Haegel}
\affil{Department of Physics, Fairfield University, Fairfield, CT
  06824} 

\altaffiltext{1}{email: kgordon@as.arizona.edu}

\begin{abstract} 
We describe the data reduction algorithms for the Multiband Imaging
Photometer for Spitzer (MIPS) instrument.  These algorithms were based
on extensive preflight testing and modeling of the Si:As (24~\micron)
and Ge:Ga (70 and 160~\micron) arrays in MIPS and have been refined
based on initial flight data. The behaviors we describe are typical of
state-of-the-art infrared focal planes operated in the low backgrounds
of space.  The Ge arrays are bulk photoconductors and therefore show a
variety of artifacts that must be removed to calibrate the data.  The
Si array, while better behaved than the Ge arrays, does show a handful
of artifacts that also must be removed to calibrate the data.  The
data reduction to remove these effects is divided into three parts.
The first part converts the non-destructively read data ramps into
slopes while removing artifacts with time constants of the order of
the exposure time.  The second part calibrates the slope measurements
while removing artifacts with time constants longer than the exposure
time.  The third part uses the redundancy inherit in the MIPS
observing modes to improve the artifact removal iteratively.  For each
of these steps, we illustrate the relevant laboratory experiments or
theoretical arguments along with the mathematical approaches taken to
calibrate the data.  Finally, we describe how these preflight
algorithms have performed on actual flight data.
\end{abstract}


\section{Introduction}
\label{sec_intro}

Most of our knowledge of the Universe at far infrared wavelengths has
been obtained with photoconductive detectors, particularly as used in
the Infrared Astronomy Satellite (IRAS) and the Infrared Space
Observatory (ISO). These detectors have been selected because they
provide excellent performance at relatively elevated operating
temperatures (compared with those needed to suppress thermal noise in
bolometers).  Similar considerations led to development of high
performance photoconductor arrays for the Multiband Imaging Photometer
for Spitzer (MIPS), namely a Ge:Ga array and a stressed Ge:Ga array
operating at 70 and 160~\micron\ respectively which were built at the
University of Arizona. To provide complementary measurements at
24~\micron, the instrument also includes a Si:As Blocked Impurity Band
(BIB) array, built at Boeing North America (BNA) under contract to the
Infrared Spectrograph (IRS) team.

In the BIB or Impurity Band Conduction (IBC) architecture, the high
impedance required to minimize Johnson noise is provided by a thin,
high-purity layer of silicon. The infrared absorption occurs in a
second layer, which can be relatively strongly doped.  Due to the 
separation of these two functions, the detector layers can be
optimized separately.  Thus, these devices can be designed
and built to have fast response, high resistance to cosmic ray
irradiation induced responsivity shifts, high quantum efficiency, and
good photometric behavior.  Because the processing necessary for high
performance silicon IBC devices has only relatively recently become
possible there is relatively little experience with them in space
astronomy missions.  An early-generation detector array was used in
the Short Wavelength Spectrometer (SWS) in ISO \citep{kes96}.
Initially, the device showed degradation due to damage by large
ionizing particle exposures when the satellite passed through the
trapped radiation belts.  Once the operating conditions were adjusted
to minimize these effects, the SWS detectors showed the expected
virtues of this type of device even though they did not achieve their
preflight sensitivity expectations \citep{deg96,her00,val00}.

At wavelengths longer than 40~\micron, photoconductors are typically
built in germanium because of the availability of impurity levels in
this material that are much more shallow than those in silicon.
Achieving the appropriate structure and simultaneously the stringent
impurity control for germanium IBC devices has proven difficult.  As a
result, all far infrared space astronomy missions have used simple
bulk photoconductors.  MIPS uses bulk gallium doped germanium (Ge:Ga)
detectors, both unstressed and stressed.  In such detectors, the same
volume of material determines both the electrical and photo-absorptive
properties, making the optimization less flexible than with Si IBC
detectors.  Consequently, their behavior retains some undesirable
properties that can be circumvented through the more complex
architecture of IBC devices.  Nonetheless, generally satisfactory
performance is possible and has been achieved in past space astronomy
missions.

The 60 and 100~\micron\ bands in IRAS \citep{neu84} utilized 15 Ge:Ga
photoconductors each. The detectors were read out with transimpedance
amplifiers that used junction field effect transistor (JFET) first
stages mounted in a way that isolated them thermally.  This allowed
these transistors to be heated, resulting in low noise and stable operation
\citep{rie81, low84}. Detector calibration was maintained by flashing
reverse bolometer stimulators mounted in the center of the telescope
secondary mirror, and cosmic ray effects were erased by boosting the
detector bias to breakdown \citep{bei88}. The intrinsic performance of
the detectors was limited by the Johnson noise of the transimpedance
amplifier (TIA) feedback resistors and by other noise sources
associated with the readout. The in-flight performance was similar to
expectations from pre-flight calibrations.

The ISOPHOT instrument \citep{lem96} carried a $3\times 3$ array of
unstressed Ge:Ga detectors operating from 50 to 105~\micron\ and a
$2\times 2$ array of stressed devices operating from 120 to
200~\micron. The readout was by a capacitive transimpedance metal
oxide semiconductor field-effect transistor (MOSFET) amplifier whose
processing had been adjusted to improve its performance at low
temperatures \citep{die92}.  Calibration was assisted with a
stimulator built into the instrument, which could be viewed by
adjusting the position of a scan mirror.  In practice, the unstressed
focal plane never achieved the performance level anticipated from
laboratory measurements of its noise equivalent power (NEP). The
performance of the stressed devices was substantially better, due in
part to the relatively large fast response component of these devices
(compared with the slow component) and their better thermal isolation
from the readout amplifiers.

The Long Wavelength Spectrometer (LWS) instrument on ISO \citep{swi96}
used a single Ge:Be detector, five Ge:Ga detectors, and four stressed
Ge:Ga detectors. The readouts were based on JFETs, mounted with
thermal isolation and heated to a temperature where they operated with
good stability and low noise. The readout circuit was an integrating
source follower and NEPs of $\sim 1\times 10^{-18}~W/Hz^{1/2}$ were
measured in the laboratory \citep{chu93}.  Calibration was assisted
with built in stimulators that were flashed between spectral scans.
On orbit, it was found that frequent small glitches, probably
associated with cosmic ray hits, limited the maximum integration times
to shorter values than had been anticipated and also required a lower
operating voltage \citep{bur98, swi00}.  With these mitigations, 
the NEPs were found to be $\sim$4 higher in orbit than expected from
ground test data \citep{swi00}.

In MIPS, the Ge:Ga detectors are carefully isolated thermally from
their readouts and operated at sufficiently cold temperatures that
their dark currents are low and stable. The MOSFET-based readouts use
a specialized foundry process that provides them with good DC
stability even at the low operating temperature of $\sim$1.5K.  This
feature, combined with the capacitive TIA (CTIA) circuit, maintains
the detector bias accurately.  A scan mirror (based on a design
provided by T.\ de Graauw) modulates the signals on a pixel so
measurements can be obtained from the relatively well-behaved
\citep{hae01} fast component of the detector response. Responsivity 
variations are tracked with the aid of frequent stimulator
flashes.  Finally, the instrument operations force observers to combine
many short observations of a source into a single measurement. The
high level of redundancy in the data helps identify outlier
signals and also improves the calibration by simple averaging over
variations. The efficacy of this operational approach is confirmed by
the on-orbit results.  Details 
on the design and construction of MIPS can be
found in \citet{hei98, sch98, you98}.  The inflight performance
of MIPS is described by \citet{rie04}.

This paper describes the approaches for reduction and
calibration of the MIPS data.  Section~\ref{sec_challenge} details the
challenges of using Si and Ge detectors in a space astronomy mission.
Section~\ref{sec_design} gives a summary of the design and operational
features of MIPS that address these challenges.
Section~\ref{sec_overview} gives an overview of the three stages of
MIPS data reduction.  These stages are discussed in more detail in the
following three sections.  Section~\ref{sec_ramps_to_slopes} details
the processing steps to turn the integration ramps into measured
slopes.  Section~\ref{sec_slope_cal} discusses the corrections to
transform the slopes into calibrated fluxes.  Section~\ref{sec_redund}
gives a brief overview of the use of the inherent
redundancy in the observations to further improve the reduction.
Section~\ref{sec_inflight} gives the results of initial testing of
these reduction techniques with flight data.  Finally,
Section~\ref{sec_summary} provides a summary. 

\section{The Challenge}
 \label{sec_challenge}

\subsection{Germanium Arrays (70 \& 160 \micron)}

At high backgrounds, such as might be encountered in an airborne
instrument, far infrared photoconductors behave relatively well, with
rapid adjustment of the detector resistance appropriate
to a change in illumination level. As the background is decreased, the
adjustment to equilibrium levels occurs in a multistep process with
multiple associated time constants as discussed below. Thus, the
detectors can be used in a straightforward manner at high backgrounds
but precautions must be taken at low ones to track the calibration.
For a more detailed discussion see \citet{rie02}.

The fast response component in these detectors results from the
current conducted within the detector volume associated with the drift
of charge carriers freed by absorption of photons.  The speed of this
component is controlled by the propagation of a zone boundary with
drift velocity $v_d$, so that the time constant is given by the free
carrier lifetime divided by the photoconductive gain.  This time is
very fast (microseconds or shorter) in comparison to normal detection
standards.  However, as charge moves within the detector, the
electrical equilibrium must be maintained. For example, charge
carriers generated by photoionization are removed from the detector
when they drift to a contact. They are replaced by injection of new
charge carriers from the opposite contact, but the necessity for new
charge can only be conveyed across the detector at a characteristic
time proportional to the ``dielectric relaxation time'', basically its
capacitive or $RC$ time constant:
\begin{equation}
\tau_d = \frac{\kappa_0 \epsilon_0}{\mu n_0 q}.
\end{equation}
Here, $\kappa_0$ is the dielectric constant of the material and $\mu$
is the mobility for the charge carrier of interest, $\epsilon_0$ is
the permittivity of free space, $n_0$ is the density of free carriers,
and $q$ is the charge of the electron. 

The slow response components arise from this phenomenon. The form of
this time constant makes explicit the dependence on illumination level
through the density of free charge carriers, $n_o$. In fully
illuminated detectors (for example, the integrating cavities used for
the 160~\micron\ array) and at the low backgrounds appropriate for
space-borne operation, $\tau_d$ can be tens of seconds.  In transverse
contact detectors, such as those used for the MIPS 70~\micron\ array,
the part of the detector volume near the injecting contact may be
poorly illuminated and have large resistance. The detector therefore
adjusts to a new equilibrium only at the large dielectric time
constant of this layer, which can be hundreds of seconds at low
backgrounds. The initial shift of charge in the detector can set up a
space charge that reduces the field in the bulk of the device, leading
to a reduction of responsivity following the initial fast
response. From its appearance on a plot of response versus time, this
behavior is described as ``hook'' response. As the field is restored
at a characteristic rate of $\tau_d$, the response grows slowly to a
new equilibrium value.  See \citet{hae01} for detailed modeling of
these effects.

In space applications, ionizing particles such as cosmic rays also
affect the calibration of these detectors. The electrons freed by a
cosmic ray hit can be captured by ionized minority impurities,
reducing the effective compensation and increasing the
responsivity. The shifts in detector characteristics can be removed by
warming it to a temperature that re-establishes thermal equilibrium,
and then cooling it back to proper operating conditions. Between such
anneal cycles, the responsivity needs to be tracked to yield
calibrated data.  All successful uses of far infrared photoconductors
at low backgrounds have included local relative calibrators of reverse
bolometer design that allow an accurately repeatable amount of light
to be put on the detector. These stimulators allow frequent
measurement of the relative detector responsivity.  In general, this
strategy is most successful when the conditions of measurement are
changed the least to carry out the relative calibration. The MIPS
instrument includes such calibrators, which are flashed approximately
every two minutes. Based upon data obtained at a proton accelerator
and in space, the average increase in response over a two minute
period in the space environment can be 0.5\% to 1\%, so the
calibration interval allows tracking the response accurately.

\subsection{Silicon Array (24~\micron)}

Although the detectors in the silicon array are expected to perform
well photometrically, the array as a system shows a number of effects
that must be removed to obtain calibrated data. The array is operated
well below the freezeout temperature for the dopants in the silicon
readout (the readout circuit uses a different foundry process from
that developed for the Ge detectors).  Therefore, the array must be
operated in a continuous read mode to avoid setting up drifts in the
outputs that would degrade the read noise. The flight electronics and
software are designed to maintain a steady read rate of once per half
MIPS second (see \S\ref{sec_data_collection}). When the array is first
turned on, the transient effects of the readout cause a slow drift in
the outputs.  Much of this effect can be removed by annealing the
array, which is the standard procedure for starting the MIPS
24~\micron\ operations.

The array shows an effect termed ``droop.'' The output of the device
is proportional to the signal it has collected, plus a second term
that is proportional to the average signal over the entire array.  In
addition, the 24~\micron\ array has a number of smaller effects (e.g.,
rowdroop, electronic nonlinearities, etc.) which are described later
in this paper.

\section{Design and Operation of MIPS}
\label{sec_design}

The design and operation of MIPS is summarized here, paying special
attention to those areas which answer the challenges outlined above
and, therefore, produce data that can be reduced successfully.

\subsection{Instrument Overview}

MIPS has three instrument sections, one for 24~\micron\ imaging, one
for 70~\micron\ imaging and low resolution spectroscopy, and one for
160~\micron\ imaging.  Light is directed into the three sections off a
single axis scan mirror.

The 24~\micron\ section uses a $128\times 128$ pixel Si:As IBC array
and operates in a fixed broad spectral band extending from 21 to about
27~\micron\ (the long wavelength cutoff is determined by the
photo-absorptive cutoff of the detector array).  After light enters
this arm of the instrument from a pickoff mirror, it is brought to a
pupil on a facet of the scan mirror.  It is reflected off this mirror
into imaging optics that relay the telescope focal plane to the
detector array at a scale of $2\farcs 5$ per pixel corresponding to a
$\lambda/2.2D$ sampling of the point spread function, where $D$ is the
telescope aperture.  The surface area of a 24~\micron\ pixel is $75 \times
75$ $\micron^2$.  The field of view provided by this array is
$5\farcm 3$.  A reverse bolometer stimulator in this optical train
allows relative calibration signals to be projected onto the array.
The scan mirror allows images to be dithered on the array without the
overheads associated with moving and stabilizing the spacecraft.  It
also enables an efficient mode of mapping (scan mapping) in which the
spacecraft is scanned slowly across the sky and the scan mirror is
driven in a sawtooth waveform that counters the spacecraft motion,
freezing the images on the detector array during integrations.

The 70~\micron\ section uses a $32\times 32$ pixel Ge:Ga array
sensitive from 53 to 107~\micron.  A cable failure external to the
instrument has disabled half of the array and the following
description reflects this situation.  The light from the telescope is
reflected into the instrument off a second pickoff mirror. It is
brought to a pupil at a second facet of the scan mirror and from there
passes through optics that bring it to the detector array.  For this
arm of the instrument, there are actually three optical trains that
can relay the light to the array; the scan mirror is used to select
the path to be used for an observation.  One train provides imaging
over a field $2\farcm 7 \times 5\farcm 3$, with a pixel scale of $9\farcs 8$
corresponding to a $\lambda/1.8D$ sampling of the point spread
function.  The physical size of a 70~\micron\ pixel is $0.75 \times
0.75$ mm$^2$ and 2~mm long in the direction of the optical axis.  This
train provides imaging over a fixed photometric band
from 55 to 86~\micron. The scan mirror feeds this mode when it is in
position to feed the other two arrays, so imaging can be done on all
three arrays simultaneously.  A second train also provides imaging in
the same band, but with the focal plane magnified by a factor of two
to $4\farcs 9$ per pixel.  This mode is provided for imaging compact
sources where the maximum possible angular resolution is desired: the
pixel scale corresponds to $\lambda/3.5D$ at the center wavelength of
the filter band.  The third train
brings the light into a spectrometer, with spectral resolution of $R =
\lambda / \Delta \lambda \sim 25 - 15$ from $53 - 107$~\micron.  In
this Spectral Energy Distribution (SED) instrument mode, light is
directed to a reflective ``slit'' and then to a concave reflective
diffraction grating that disperses the light and images the spectrum
onto a portion of the 70~\micron\ array.  The slit is 16~pixels long
and 2~pixels wide, corresponding to $2\farcm 7 \times 0\farcm 32$ on
the sky.  The dispersion is 1.73~\micron\ pixel$^{-1}$.  Reverse
bolometer stimulators are 
provided for calibration, and the scan mirror provides the dithered
and scan mapping modes of operation at 70~\micron\ as have been
described for the 24~\micron\ array.

The 160~\micron\ section shares the pickoff mirror and scan mirror
facet with the 24~\micron\ band.  After the light has been reflected
off the scan mirror, the telescope focal plane is reimaged and
divided, with part going to the Si:As array and part going to a
stressed Ge:Ga array, operating in a fixed filter band from 140 to
180~\micron.  This array has $2\times 20$ pixels, arranged to provide
an imaging field $5\farcm 3$ long in the direction orthogonal to the
scan mirror motion with the two rows of detectors spaced
such that there is a gap one pixel wide between the two rows.  This
pixel size provides $\lambda/2.2D$ sampling of the point spread
function.  The physical size of a 160~\micron\ pixel is $0.81 \times
0.81 \times 0.81$ mm$^3$.
Reverse bolometer stimulators are included in the optical 
train, and the scan mirror provides modes similar to those with the
other two arrays.

\subsection{Stimulators}
\label{sec_stims}

A key aspect of the calibration of the MIPS Ge arrays is the frequent
use of stimulators \citep{bee02} to track responsivity variations.
The emitters in these devices are sapphire plates blackened with a
thin deposition of bismuth, which also acts as an electrical
resistor. The emitters are suspended in a metal ring by nylon supports
and their electrical leads.  When a controlled current is run through
the device, the sapphire plate is rapidly heated by ohmic losses in
the metallized layer.  The thermal emission is used to track changes
in detector response in a relative manner; hence these devices are
described as stimulators rather than calibrators.  Because of the
large responsivity of the detector arrays, it is necessary to operate
these stimulators highly inefficiently to ensure accurate control
without blinding the detectors.  They are mounted inside cavities that are
intentionally designed to be inefficient (e.g., black walls, small
exit holes).  This allows the stimulators to be run
at high enough voltage to be stable and emit at a reasonable effective
temperature.

\begin{figure*}[tp]
\plottwo{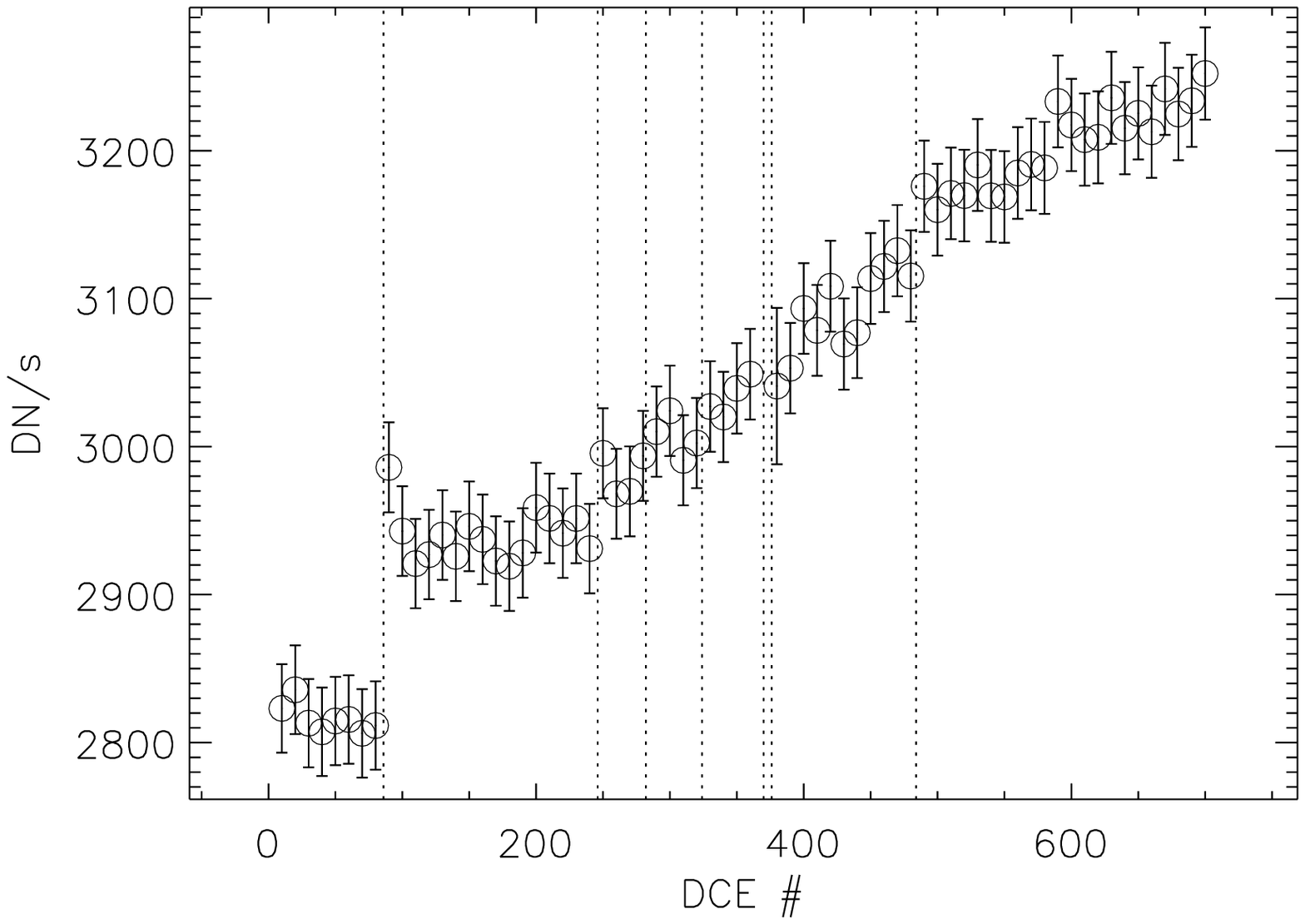}{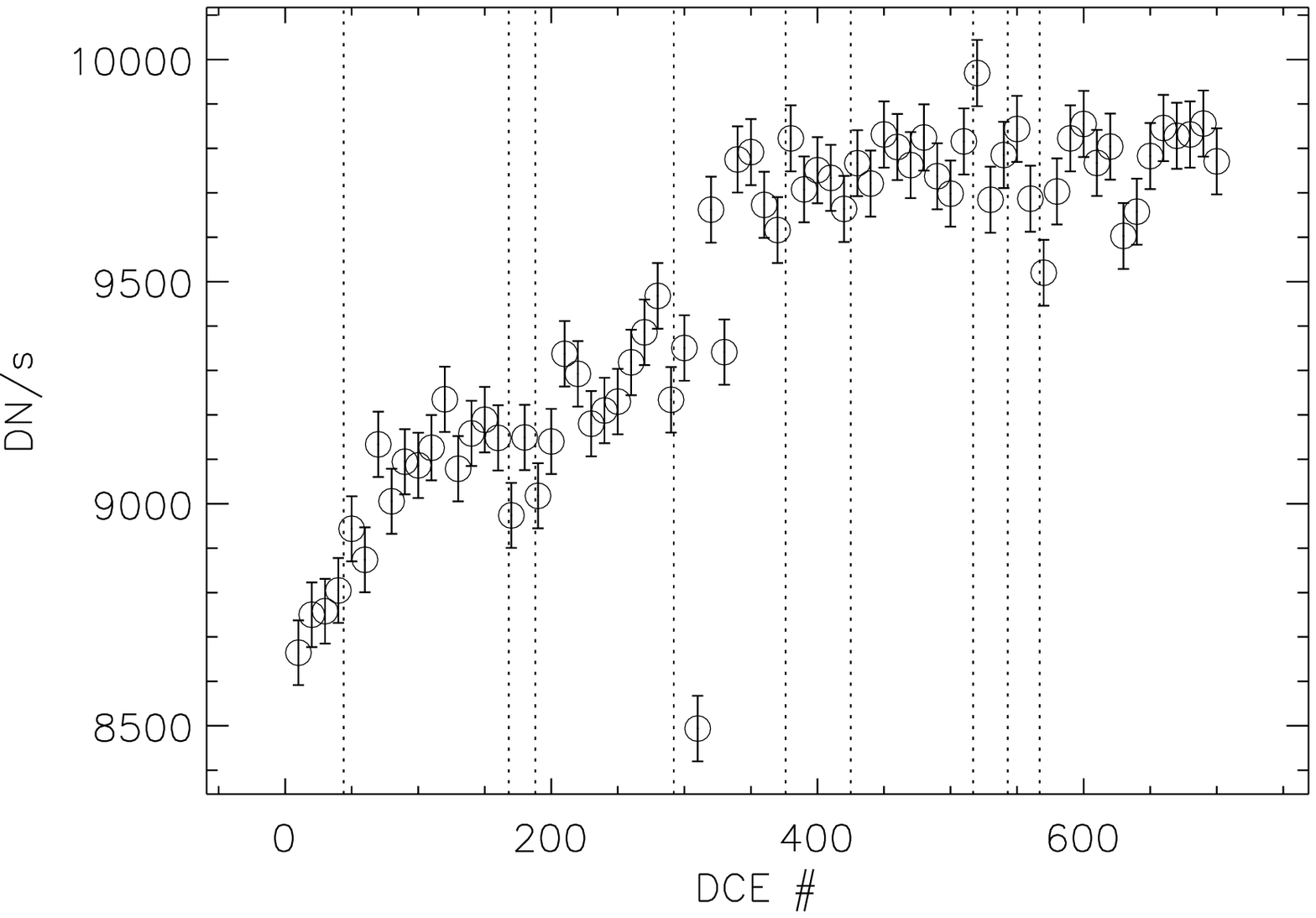}
\caption{The stim flash amplitudes for a single pixel of the
70~\micron\ (left) and 160~\micron\ (right) arrays are plotted from a
2 hour test where the stim was flashed approximately every 2 minutes
on a constant background.  The vertical dotted lines denote the image in
which a cosmic ray was detected.  On the 70~\micron\ plot, the first
cosmic ray can be seen to have caused a $\sim$3\% responsivity
increase.  The unit DN/s refers to Data Numbers per second.  The
x-axis gives the number of images (DCE is defined in
\S\ref{sec_data_collection}) taken where each image 
represents about 11 seconds of data.  \label{fig_stim_repeat}} 
\end{figure*}

The constant-amplitude stimulator flashes provide a means of tracking
the responsivity drift inherent in the Ge detectors.
Figure~\ref{fig_stim_repeat} illustrates the importance of tracking
the responsivity variations of Ge detectors with as fine a time
resolution as feasible.  The repeatability of the stimulator
measurements is a function of both the background seen by the detector
element as well as the amplitude of the stimulator (stim) signal above
the background.  The repeatability of a measurement of the stim signal
improves with decreasing background and increasing stim amplitude.
For both the 70 and 160~\micron\ arrays, in ground testing stim
amplitudes of greater than $\sim$7500~DN/s above the background
yielded a repeatability of better than $\sim$1\%\ on most
backgrounds. Setting the stim amplitudes at this level provides a
balance between repeatability of the stims and the range of
backgrounds accessible to observation without saturation.  At this
level, well over 95\%\ of the sky should be observable without
saturating stim flash measurements at both 70 and 160~\micron.

Additional complications at 160~\micron\ include a strong illumination
gradient in the stim flash illumination pattern from one end of the
array to the other as well as a large increase in the responsivity of
the array with exposure to cosmic rays.  It is not possible to set the
stim amplitude at the optimum 7500~DN/s across the whole array due to
a factor of four gradient in the stim amplitude across the array.  The
on-orbit stim amplitude was set to provide an optimal amplitude over
the majority of the 160~\micron\ pixels.  The degradation in stim
repeatability on the low illumination region can be mitigated by an
observing strategy that dithers the image such that the same region of
the sky spends equal amounts of time on both regions of the detector.

\subsection{Anneals}
\label{sec_anneals}

Both of the Ge arrays show calibration shifts with
even small exposure to ionizing radiation.  The effects of
ionizing particles were tested using characterization arrays (see
\S\ref{sec_lab_test}) at the
University of California, Davis accelerator.  The proton beam was
attenuated to reduce the particle impact rate to a level similar to
that expected on orbit.  The energy of the particles was such that
each impact was strongly ionizing, depositing much more energy in the
detector volume than is expected from a typical cosmic ray.  Thus,
these tests served as a worst-case model of the detector response to
cosmic-rays on orbit.

The detector responsivity slowly increased with time under exposure to
the proton beam.  The rate of responsivity increase on the 70~\micron\
array was comparable to that observed under typical illumination
conditions (cf.\ Fig.~\ref{fig_stim_repeat}) without the proton beam,
suggesting that accumulated transient response from the background and
signals inside the cold test chamber and the photon flux at the
accelerator contribute similarly to the 
responsivity increase.  If the particle impacts at the accelerator
really represent a worst-case scenario, this suggests that the
on-orbit responsivity increase of the 70~\micron\ array may be
dominated by photon flux rather than cosmic ray effects. In contrast,
the 160~\micron\ array showed a large responsivity increase with
increasing radiation dose.

If they are of modest size, such responsivity shifts can be determined
and removed during calibration through use of the stimulator
observations.  However, when the shifts are large, they are also
highly unstable and can result in substantial excess noise. Three
methods were tested to remove such effects: re-thermalization of the
detectors by heating them (anneals), exposing the detectors to a
bright photon source, and boosting the detector bias above
breakdown. Our experiments indicated that the latter two methods
produced little benefit.  Although the instrument design permits use
of all three techniques, we remove radiation damage to the
Ge arrays by periodically thermally annealing the detectors.

\subsection{Observing Modes}

\begin{figure*}[tp]
\plotone{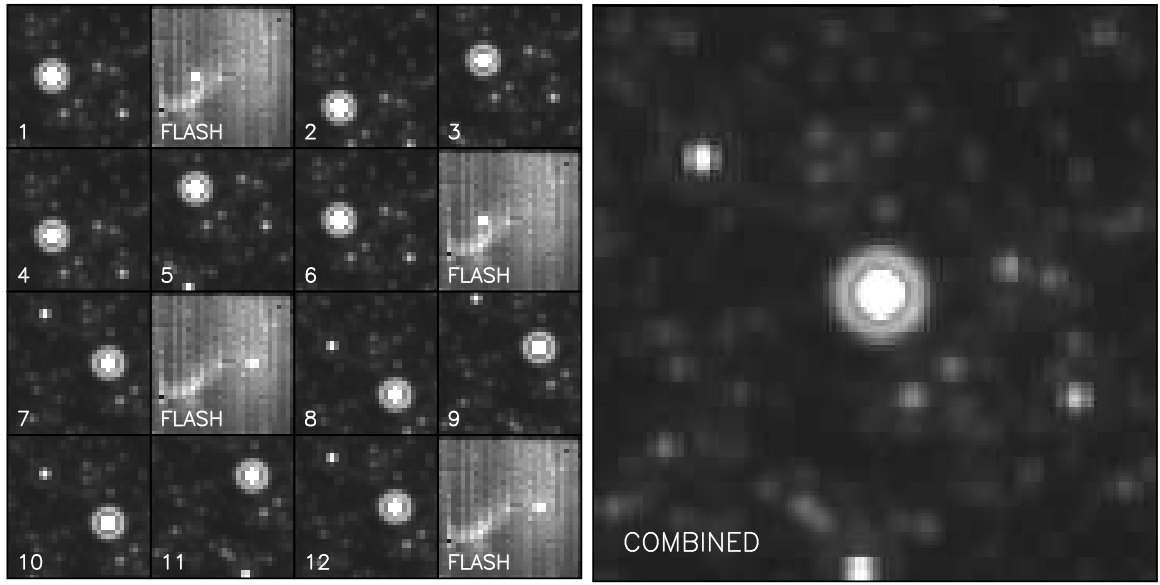}
\caption{The images for a single observation cycle in 70~\micron\ Compact Source
Photometry mode.  The numeral on each individual image gives the
object image number in the cycle.  The ``FLASH'' designation
corresponds to a stim flash image and the image taken before the stim
flash gives the background on top of which the stim is flashed.  The
``COMBINED'' image was made with pixels 1/4 the original size.  Note
that an object in the center of the combined image is significantly
better sampled than one near the edges.
\label{fig_phot70}}
\end{figure*}

\begin{figure*}[tp]
\plotone{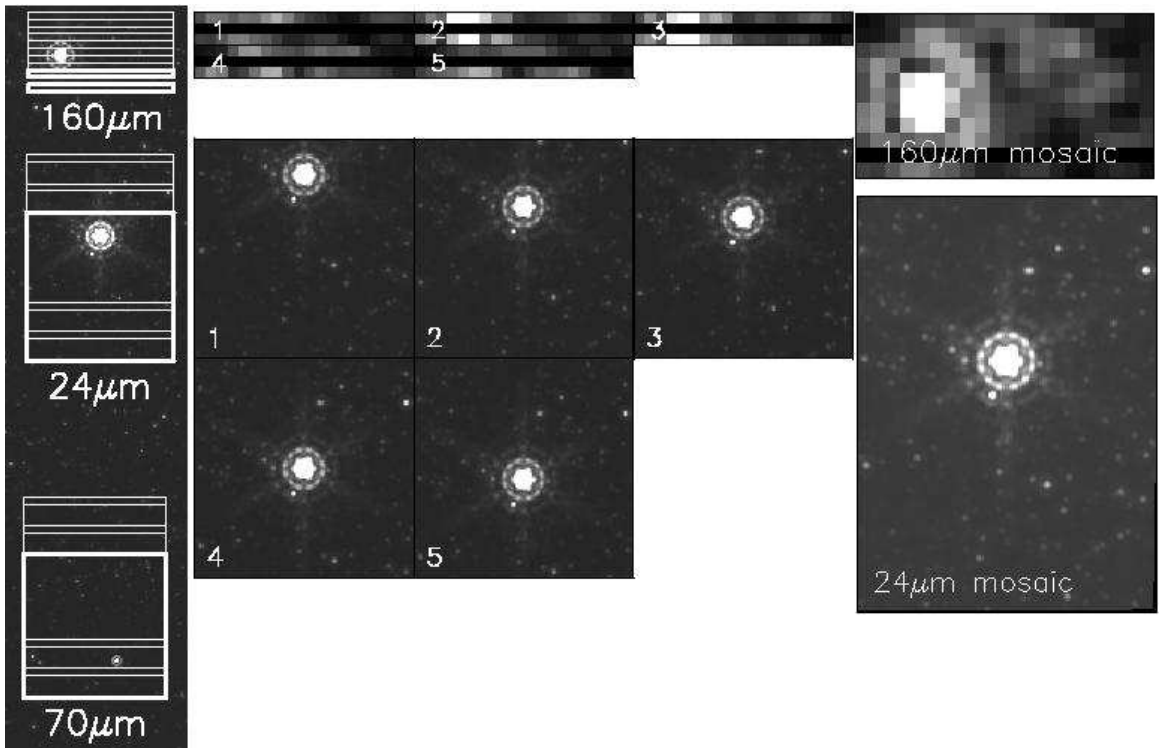}
\caption{A visualization for the Scan Map mode. The stim flashes are
not shown.  On the left, the locations of 5 simultaneous images at 24,
70, and 160~\micron\ are shown on a sky image at the 24~\micron\
resolution.  The single field of view for each array is denoted by a
bold outline.  The middle shows the individual images at 24 and
160~\micron; the 70~\micron\ images resemble the 24~\micron\ images,
just offset downward.  On the right, the 24 and 160~\micron\ mosaics
created from the 5 individual frames are shown.
\label{fig_scanmode}}  
\end{figure*}

There are four MIPS observing modes, all of which have been designed
to provide a high level of redundancy to ensure good quality data
(especially for the Ge arrays).  The Photometry mode is for point and
small sources.  As an example of the redundancy inherit in MIPS
observations, a visualization of a single Photometry Mode cycle is
shown in Fig.~\ref{fig_phot70} for 70~\micron.  The Scan Map Mode
provides efficient, simultaneous mapping at 24, 70, and 160~\micron\
by using a ramp motion for the scan mirror to compensate for
continuous telescope motion, effectively freezing images of the sky on
the arrays.  A visualization of a small portion of a scan leg is shown
in Fig.~\ref{fig_scanmode}.  The SED mode provides 53 to 107~\micron\
spectra with a resolution R $\approx 25 - 15$.  The Total Power Mode
(TPM) is for making absolute measurements of extended emissions.

\subsection{Data Collection}
 \label{sec_data_collection}

The pacing of the MIPS data collection is based on a ``MIPS second.''
A MIPS second is approximately 1.049 seconds, and has been selected to
synchronize the data collection with potential sources of periodic
noise, such as the computer clock or the oscillators in the power
supplies. To first order, this design prevents the down-conversion of
pickup from these potential noise sources into the astronomical
signals. The data are taken in Data Collection Events (DCEs); at the
end of a DCE, the array is reset before taking more data.  DCEs are
currently limited to 3, 4, 10, or 30 MIPS seconds for the 24~\micron\
array and 3, 10, or 10 MIPS seconds for the 70 and 160~\micron\
arrays.

During a DCE, each pixel generates a voltage ramp on the array output,
as the charge from incoming photons is accumulated on the input node
of its integrating amplifier.  These ramps are the basic data
collected by all three arrays.  The 24~\micron\ array is
non-destructively read out every 1/2 MIPS second while the 70 and
160~\micron\ arrays are non-destructively read out every 1/8 MIPS
second. All the samples are downlinked for the 70 and 160~\micron\
arrays, but this is not possible for the 24~\micron\ array due to
bandwidth restrictions.  The 24~\micron\ array has two data modes, SUR
and RAW.  Most 24~\micron\ data are taken in SUR mode in which the
ramps are fitted to a line, and only the fitted slope and first
difference (the difference between the first two reads in the ramp) are
downlinked.  The RAW mode downlinks the full 24~\micron\ ramps, but
this mode is used only for engineering observations.

\section{Overview of MIPS Data Processing}
\label{sec_overview}

There are three natural steps in reducing data from integrating
amplifiers: (1) converting the integration ramps to slopes; (2)
further time-domain processing of the slope images; and (3) processing
of dithered images in the spatial domain.  For detectors that do not
have time-dependent responsivities, only the first and last steps are
usually important.  This is strongly not the case for the MIPS Ge
arrays and also mildly not so for the MIPS Si array.

As a result, MIPS processing includes all three steps. First, the
integration ramps are converted into slopes (DN/s) while removing
instrumental signatures with time constants on the order of the DCE
exposure times (\S\ref{sec_ramps_to_slopes}).  Second, the slopes are
calibrated and instrumental signatures with time constants longer than
the DCE exposure times are removed (\S\ref{sec_slope_cal}).  Third,
the redundancy inherent in the MIPS observing modes allows a second
pass at removing instrumental signatures (\S\ref{sec_redund}).  The
algorithms used in the first two steps have mostly been determined.
The main algorithms used by the third step are being optimized with
actual data taken on 
orbit.  Portions of the reduction algorithms described in this paper
were presented in a preliminary form by \citet{hes00}.

\begin{figure*}
\plotone{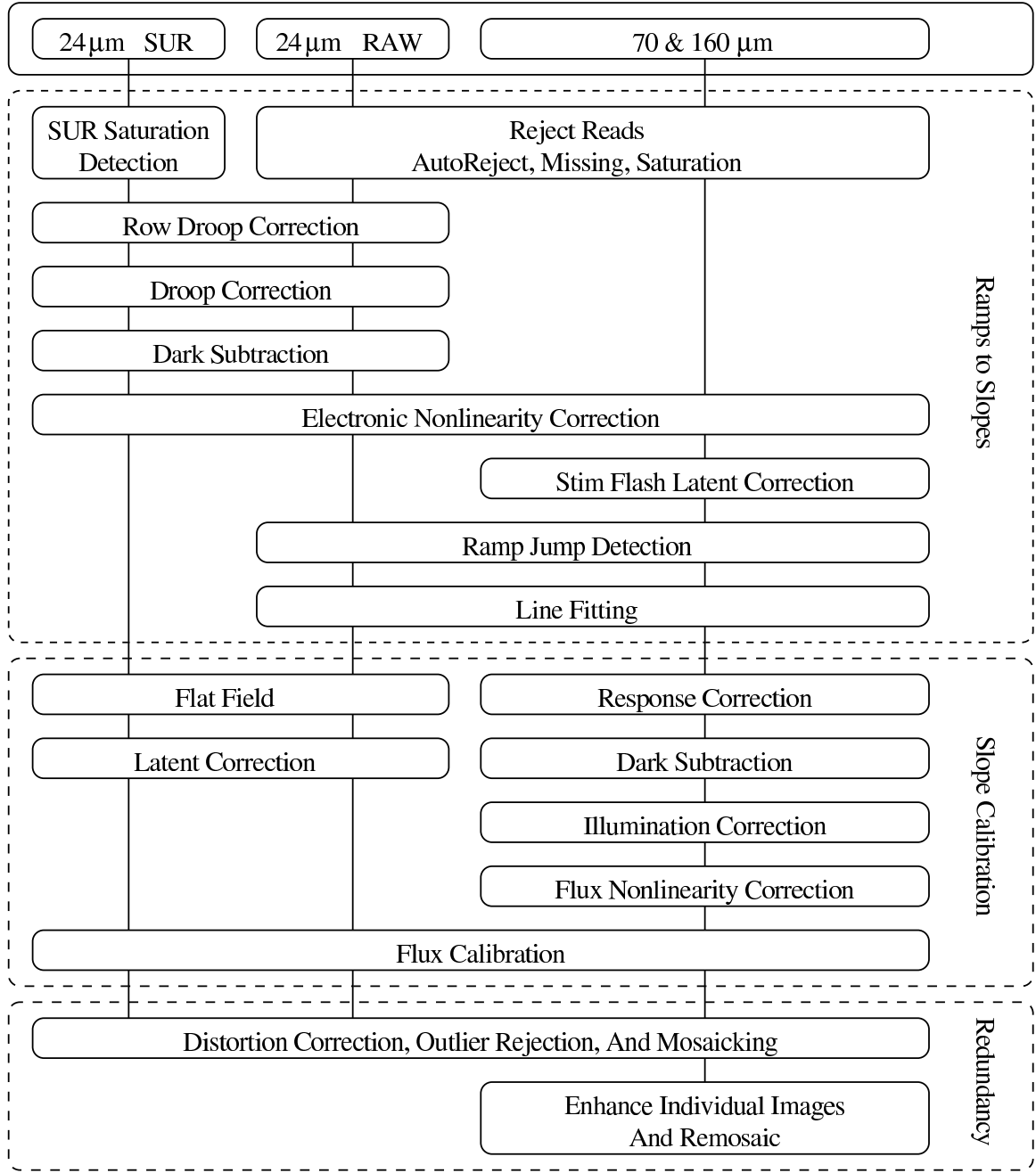}
\caption{Graphical representation of the flow of the reduction of MIPS
data.  \label{fig_flowchart}}
\end{figure*}

We made extensive use of laboratory testing and theoretical
investigations in choosing and ordering the relevant
steps. Figure~\ref{fig_flowchart} is a graphical representation of the
specific tasks in each of the three processing steps.

\subsection{Laboratory Testing of Ge Arrays \label{sec_lab_test}}

Three versions of the 70 and 160~\micron\ arrays were constructed: a
flight array, a flight spare array, and a characterization array.
Before integration into the instrument, the basic performance of the
flight and flight spare arrays was measured (i.e., read noise, dark
current, NEP, etc.).  The characterization arrays were then installed
in the two specialized dewars previously used for the flight and
flight spare array testing.  These arrays are used to determine the
detailed behaviors of the 70 and 160~\micron\ detectors.  This
knowledge was then used to design observations with the
flight arrays to remove specific
Ge detector effects.  The ability to do extensive testing on the
characterization arrays has been crucial to the development of the
data reduction algorithms for the Ge arrays detailed in this paper.

In addition to testing at the array level, testing at the instrument
level was carried out using the Low Background Test Chamber (LBTC).
The LBTC was constructed to allow for testing of the full MIPS
instrument and, thus, had a number of independently controlled
stimulators including pinhole stimulators providing point sources for
testing.  The LBTC allowed for the imaging performance of the full
instrument to be tested as well as providing for extensive testing of
the 24~\micron\ array.

Additional details of the laboratory testing can be found in
\citet{you03}. 

\subsection{Numerical Modeling of Ge Arrays}

We also carried out detailed numerical modeling of the behavior of the Ge
arrays.  This modeling allowed effects found in the laboratory testing to
be investigated in more detail.  For example, the modeling was able to
show that the difference in hook behaviors between the 70 and
160~\micron\ arrays was due to their different illuminations
\citep{hae01}.  The numerical modeling was also crucial to the
understanding of the behavior of small signals on the detectors.  For
example, this modeling was able to determine that the stim flash
latents (see \S\ref{sec_sflash_latent}) were additive, not
multiplicative.  This understanding then guided the efforts to remove
this signal.

\section{Ramps to Slopes}
\label{sec_ramps_to_slopes}

\begin{figure}
\plotone{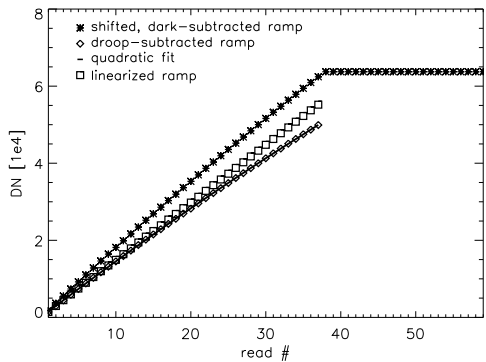}
\caption{Example of a 24~\micron\ RAW data ramp for a 30 second DCE,
showing several steps along the processing pipeline.  Asterisks
represent the ramp after dark subtraction and show ADC saturation is
reached at read 38.  Diamonds are the ramp after droop
subtraction (the saturated reads are not shown since they are not used
in subsequent processing steps).  The quadratic fit (used to derive
the electronic nonlinearity correction) to these points is shown with
the solid line.  The final, linearized ramp is shown with the open
squares.
\label{fig_24ramp}}
\end{figure}

The first part of the processing fits the ramps to produce slopes for
each DCE.  The processing for the Ge (70 and 160~\micron) and Si
(24~\micron) RAW mode data is similar, differing only in the
instrumental signatures removed.  First, reads which should be
rejected from the linear fits are identified.  Reads are rejected if
they represent missing data, autoreject reads, or saturated data.
Second, the ramps are corrected for instrumental effects.  These are
dark current (Si only), rowdroop (Si only), droop (Si only),
electronic nonlinearities, and stim flash latents (Ge only).  Third,
jumps in the ramps usually caused by cosmic rays are identified.  In
the process, reads that are abnormally noisy are identified as noise
spikes.  Finally, all the continuous segments in each ramp are fit
with lines and the resulting slopes averaged to produce the final
slope for each pixel.  An example of a 24~\micron\ ramp is given in
Fig.~\ref{fig_24ramp}.  The 70 and 160~\micron\ ramps are very similar
to the 24~\micron\ ramp, except they do not have droop.  The graphical
representation of the data processing shown in
Fig.~\ref{fig_flowchart} gives the ordering of the reduction steps.
The processing for the Si SUR mode data is necessarily different as
the ramps are fit on-board and only the slope and first difference
images are downlinked.  The following subsections will describe the Si
and Ge RAW mode processing followed by a description of the necessary
differences for the Si SUR mode processing.


\subsection{Steps Common to Si \& Ge RAW Modes}

\subsubsection{Rejected Reads - Autoreject and Saturated Reads}

There are two reasons to automatically reject (autoreject) reads; to
avoid reset 
signatures and to not use the ramps beyond 2 MIPS seconds for stim
flash DCEs.  All MIPS arrays are reset at the beginning of a ramp, and
this has been seen to leave a signature in the first few reads.  In
general, this reset signature only affects the first read.  The first
read is automatically rejected for all three arrays.  This is even
true for SUR data for which the line fit is done on-board Spitzer.  The
70 and 160~\micron\ arrays can be operated with a reset in the middle
of the DCE to improve performance.  When this mode is used, the reset
signature has been seen to last for 4 reads and these 4 reads are then
automatically rejected.  In a stim flash DCE only the first 2 MIPS
seconds of a ramp are valid.  After 2 MIPS seconds, the stim is turned
off and after 2.5 MIPS seconds, a reset is applied.

Finally, all reads that are below or above the allowed limits for the
MIPS analog-to-digital converters (ADC) (soft saturation) or saturating the
70 and 160~\micron\ readout circuits (hard saturation) are flagged as
low or high saturation, respectively.

\subsubsection{Electronic Nonlinearity Correction}

All three MIPS arrays display nonlinearities that have been traced to
the electronics.  For the 24~\micron\ array these nonlinearities are
mainly due to a gradual debiasing which occurs as charge accumulates
in each pixel during an exposure.  For the 70 and 160~\micron\ arrays,
the readout circuits have been constructed to keep the same bias
voltage across the detectors even as charge accumulates.
Nevertheless, electronic nonlinearities arise due to the simplified
CTIA circuit.

The behavior of the electronic nonlinearities was determined from
extensive ground-based testing on the flight arrays.  For the
24~\micron\ array, the functional form was characterized from RAW mode
data ramps; a typical case is shown in Figure~\ref{fig_24ramp}.  The
ramps for most of the pixels can be nearly perfectly described by
quadratic polynomial fits; the linear component of the fit gives
directly the linearized signal.  For the 70 and 160~\micron\ arrays,
the electronic nonlinearities have been shown generally to have a
quadratic shape with significant deviations.  Corrections were
tabulated as a lookup table to allow for the semi-arbitrary forms.
For the 24, 70, and 160~\micron\ arrays, the maximum nonlinearity at
full well (ADC saturation) ranges over the array from $\sim 10-15\%$,
$\sim 1-2\%$, and $\sim 0.5-1\%$, respectively.

\subsubsection{Ramp Jumps - Cosmic Rays, Readout Jumps, and Noise Spikes}

The main reason discontinuities or jumps appear in MIPS ramps is
cosmic rays.  Cosmic rays strike the Ge detectors (70
and 160~\micron\ arrays) at a rate of one per pixel per twelve
seconds.  The rate on the Si detector (24~\micron\ array) is much
lower, due to its smaller pixels.  It is also possible to 
get a ramp jump due to an anomaly we have termed a readout jump.
Ground-based testing has shown that the entire output of one of the 32
readouts (4 $\times$ 8 pixels) on the 70~\micron\ array occasionally
jumps up and then jumps back down by the same DN amount approximately
1 second later.

Jumps in the ramps are detected using a combination of two methods.
First, $(n-1)$ 2-point differences are constructed from the $n$ reads
and outliers are flagged as potential ramp jumps using an iterative
sigma clipping algorithm.  These potential jumps are tested to see if
they are noise spikes or actual ramp jumps by fitting lines to the
segments on either side of the potential jump.  If the two fitted
lines imply a jump that is smaller than the expected noise, then the
jump is actually a noise spike, not a cosmic ray or readout jump.

Second, a more sensitive test for ramp jumps is performed
\citep{hes00}.  This method works by assuming each read in a ramp
segment has a ramp jump after it and fitting lines to the resulting
two subsegments on either side.  The most significant ramp jump
in the segment implied from the two
line fits is then tested to see if it is larger than the
noise.  If so, then this read is labeled as a ramp jump.  The process
can be repeated on the subsequent ramp segments until no more jumps
are found or a preset number of iterations have been performed.  As this
second method is more sensitive than the first, but significantly more
computationally intensive, we combine the two methods to achieve the
best sensitivity to ramp jumps with the least computation time.

\begin{figure*}
\plotone{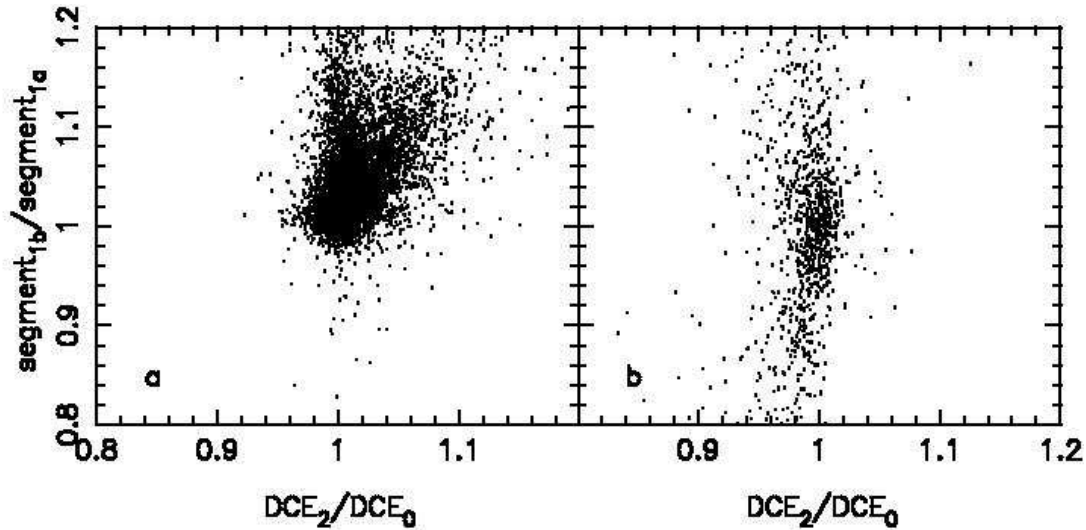}
\caption{Each panel shows the ratios of the slopes of ramp segments within
a DCE in which a particle hit occurred (designated 1a before the hit
and 1b after the hit) against the ratios of the slopes of ramps in the
DCEs before (designated DCE 0) and after (designated DCE 2) the DCE
with the hit.  The arrays were subject to a constant illumination
level and so the points should be clustered around 1 if there was no
effect due to particle hits.  The 70~\micron\ array is shown in panel
(a) while the 160~\micron\ array is shown in panel (b).
\label{figure:slope-change-after-radhit}}
\end{figure*}

We explored the signatures of cosmic rays in ramps using several
hours' worth of 70 and 160~\micron\ array data that were subject to
constant illumination.  We then extracted those ramps where we
detected ramp jumps (assumed to be due to energetic particle impacts)
and assessed the effects on the ramp after the impact.  On the
70~\micron\ array we find two main effects: a steepening of the ramp
that lasts for a few reads and a persistent
responsivity increase of $\sim 1\%$ after a big hit (see
Fig.~\ref{figure:slope-change-after-radhit}).  This is consistent with
the slow responsivity increase observed during the radiation run.
These results dictate our strategy for dealing with cosmic ray hits on
this array: several reads after a hit should be rejected from slope
fitting to ensure that the fast transient does not bias the slope
measurement, while the small responsivity increase after large hits
will be tracked by the stim flash measurements.

The 160~\micron\ array response to cosmic rays is somewhat different.
We detected no fast transient within the ramp, but the slope of the
ramp after a hit was often different from the slope before the hit.
This slope change typically did not persist into the next DCE, after a
reset had occurred, as shown in
Figure~\ref{figure:slope-change-after-radhit}.  Thus, we were unable
to detect a persistent responsivity increase due to particle impacts,
in contrast to the accelerator data (\S\ref{sec_anneals}).  Given that
we are unable to predict how the slope will change after a particle
hit and that the slope returns to its previous value after the next
reset (usually the next DCE), the conservative strategy for dealing
with particle impacts on this array is to simply ignore all data
between a particle hit and the next reset.

\subsubsection{Line Fitting}

Slopes are determined for each ramp by fitting lines to all the good
segments in a ramp.  The slope for a ramp is then the weighted average
of the slopes of the ramp segments.  The weight of each segment is
determined from the uncertainty in the segment slope as discussed in
the next paragraph.  Each good segment of a ramp is identified as
containing only good reads and not containing any ramp jumps.  Lines
are fit to these segments with the standard linear regression
algorithm.

Calculating the uncertainties on the fitted slope and zero point is
not as straightforward.  The uncertainties on each read have both a
correlated and random component.  The correlated component is due to
photon noise, as the reads are a running sum of the total number
of photons detected.  The random component is the read noise.  We have
derived equations for the linear fit uncertainties for the correlated
component following the work of \citet{spa98}.  The details of this
derivation are given in appendix~\ref{app_fit_unc}.  The slope and zero point
uncertainties are calculated for the correlated and random read
uncertainties separately and then combined in quadrature to get the
final uncertainties.

\subsection{Steps for Ge Raw Mode Only}

\subsubsection{Stim Flash Latent Correction\label{sec_sflash_latent} }

The calibration of the 70 and 160~\micron\ arrays is directly tied to
the stim flashes measured approximately every two minutes.  The
brightness of these stim flashes is set as high as possible to ensure
the best calibration (cf.\ \S\ref{sec_stims}).  These stim flashes
produce a memory effect, called a stim flash latent, that is
persistent for a brief time.  Intensive measurements
of stim flash latents have been performed at the University of Arizona
on the 70 and 160~\micron\ characterization arrays.  We determined the
time constants, amplitudes, variations with the background, and
repeatability of the stim flash latents as well as the accuracy of the
correction and the effects on the calibration of sources observed
during the latent.

To characterize the decay behavior of the latents, we fit an
exponential law to the time signal of each array pixel. Each cycle is
divided by the stim amplitude value, to have dimensionless data
(fraction signal/stim).  The function $F$ used to fit the latent is a
double exponential:
\begin{equation}
\label{eq:latent}
F(t) = b + a_1 e^{-t / \tau_1 } - a_2 e^{-t / \tau_2 }
\end{equation}
where $t$ is the time after the stim is turned off, $b$ is the
background level, $a_1$ and $a_2$ give the component amplitudes, and
$\tau_1$ and $\tau_2$ give the time constants.

\begin{figure*}[tp]
\epsscale{1.1}
\plottwo{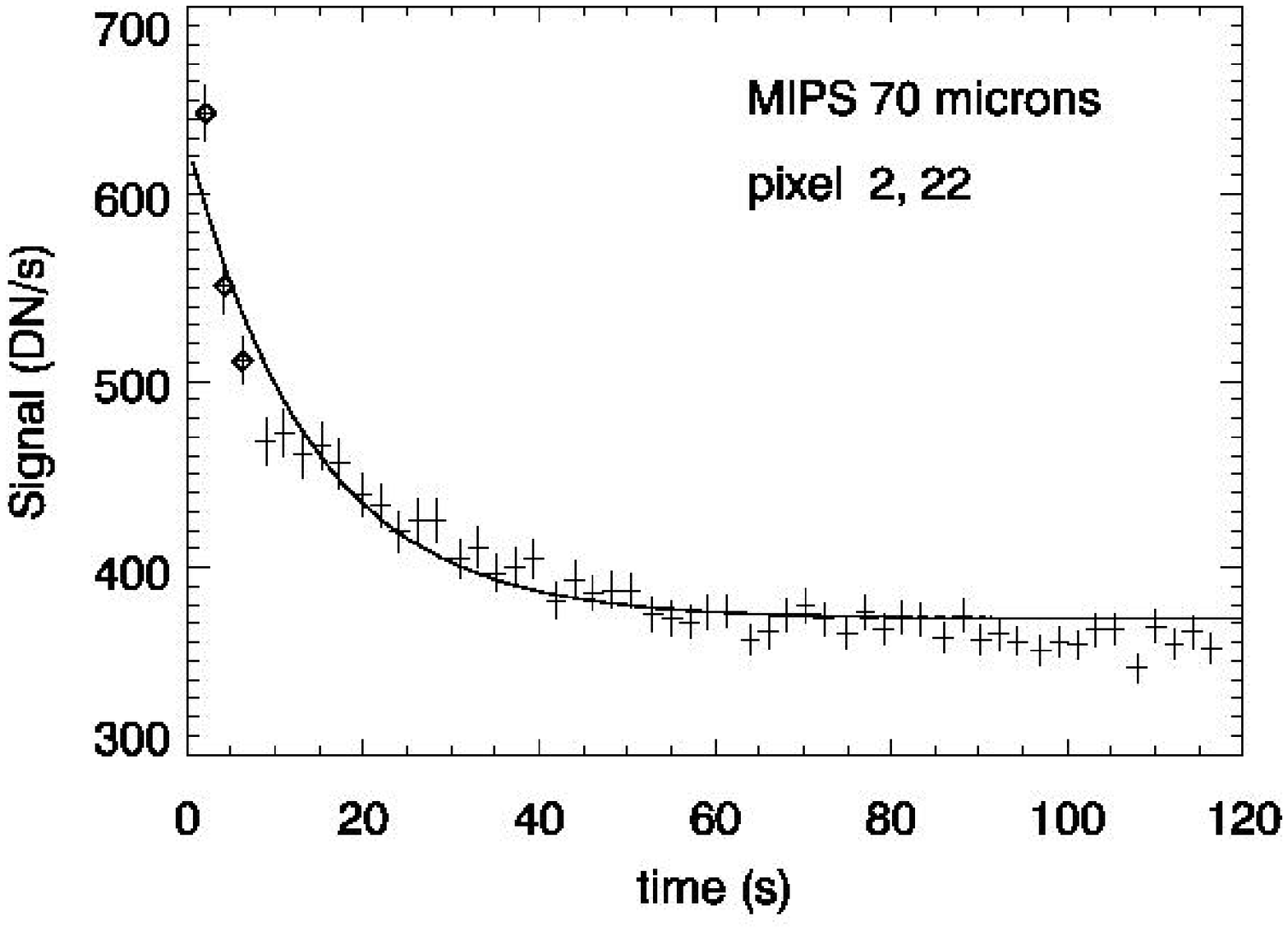}{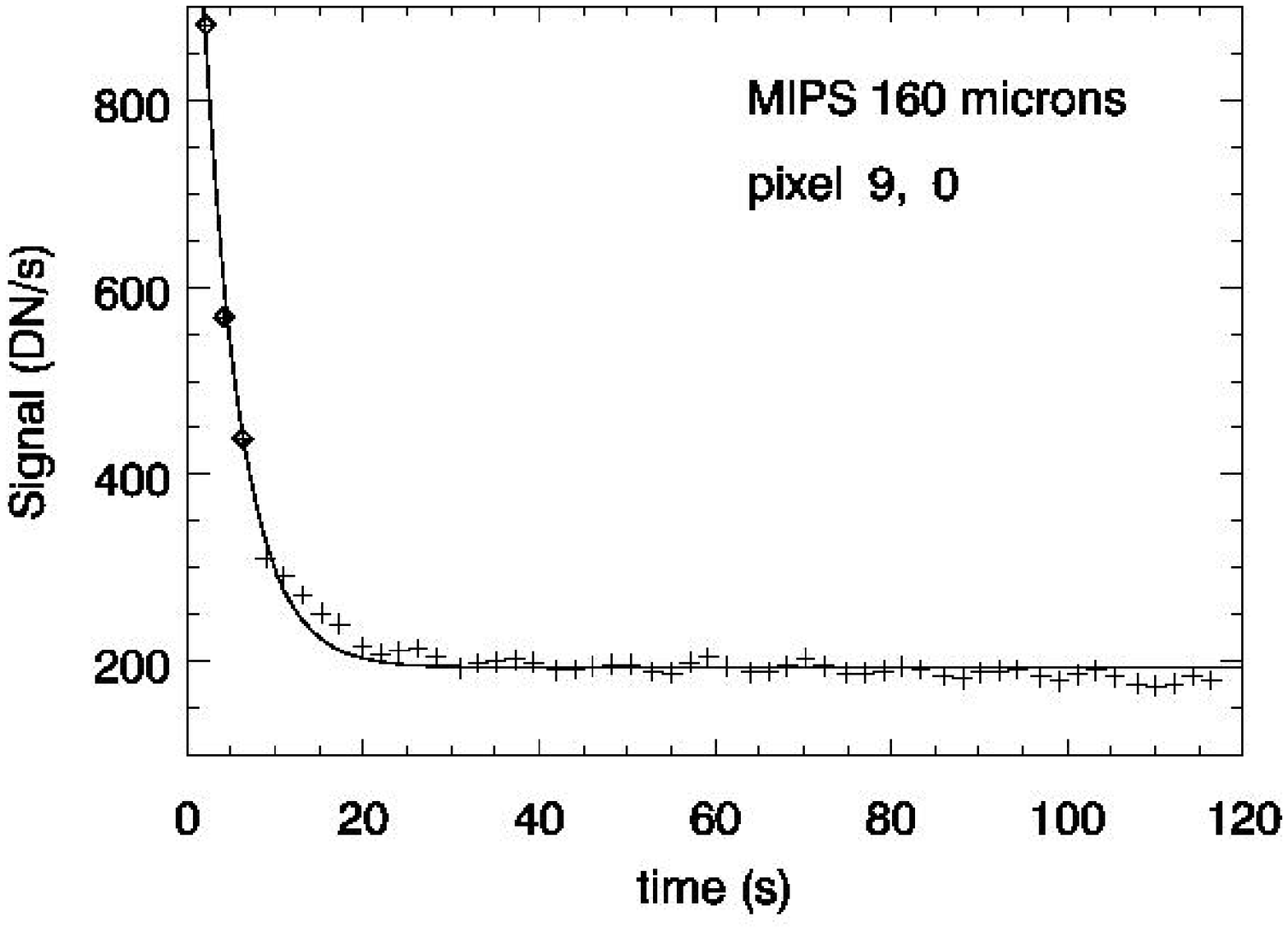}
\epsscale{1.0}
\caption{Examples of stim flash latents in the 70~\micron\ (left) and
160~\micron\ (right) characterization arrays.  These plots
represent time series of a pixel output averaged over 50 cycles.  Each
10s DCE has been subdivided in 5 sub-DCEs of 2s.  The 
diamonds give the pixel signal during the stim DCE and the crosses the
pixel signal during science DCEs.  The solid line gives the fit.  The
70~\micron\ fit parameters for pixel (2,22) are: stim flash signal =
$18004$ DN/s, $b = 372$ DN/s (2.1\% of stim flash), $a_1 = 256$ DN/s
(1.4 \% of stim flash), and $\tau_1 = 14$s.  The 160~\micron\ fit
parameters for pixel (9,0) are stim flash signal = $26985$ DN/s, $b =
193$ DN/s (0.7\% of stim flash), $a_1 = 1091$ DN/s (4.0 \% of stim
flash), and $\tau_1 = 4.3$s.
\label{fig_slatent}}
\end{figure*}

At 70~\micron, only a single exponential is needed (thus $a_2 = 0$).
The amplitude $a_1$ is always less than 3\% of the stim amplitude, and
in most of the cases below 0.5\%.  The time constant $\tau_1$ ranges
from 5s to 20s.  As a function of increasing background, $a_1$
increases and $\tau_1$ decreases.  The latents are repeatable to 15\%
or better.  An example of the stim latent of one pixel on the
70~\micron\ characterization array is given in Fig.~\ref{fig_slatent}.

At 160~\micron, the latency effect is more pronounced than at
70~\micron.  The amplitude $a_1$ is less than 5\% of the stim
amplitude.  The time constant $\tau_1$ ranges from 5 to 20s.  The
amplitude $a_2$ is less than 3\%.  The time constant $\tau_2$ equals
20s at high background, and is negligible at low background.  The
amplitude $a_1$ and time constant $\tau_1$ are almost insensitive to
the background.  The latents are repeatable to 20\% or better.
Fig.~\ref{fig_slatent} also gives an example of the stim latent of one
pixel on the 160~\micron\ characterization array.

In general, the stim flash latents are negligible $\sim$30
seconds after the stim is turned off.  In the first 30 seconds, the
calibration of a point source might be overestimated by 1\% at
70~\micron\ and 12\% at 160~\micron\ if no correction is applied.
To correct for the stim latent contribution to the pixel signal, we
apply a time-dependent correction at the ramp level. We subtract the
latent contribution, which is obtained by integrating
Eq.~\ref{eq:latent}.  On pre-flight data, the amplitude of the
latents after correction is reduced by a factor of $\sim$2 at
70~\micron\ and $\sim$4 at 160~\micron.

\subsection{Steps for Si RAW Mode Only}

\subsubsection{Rowdroop Subtraction}

\begin{figure*}
\plottwo{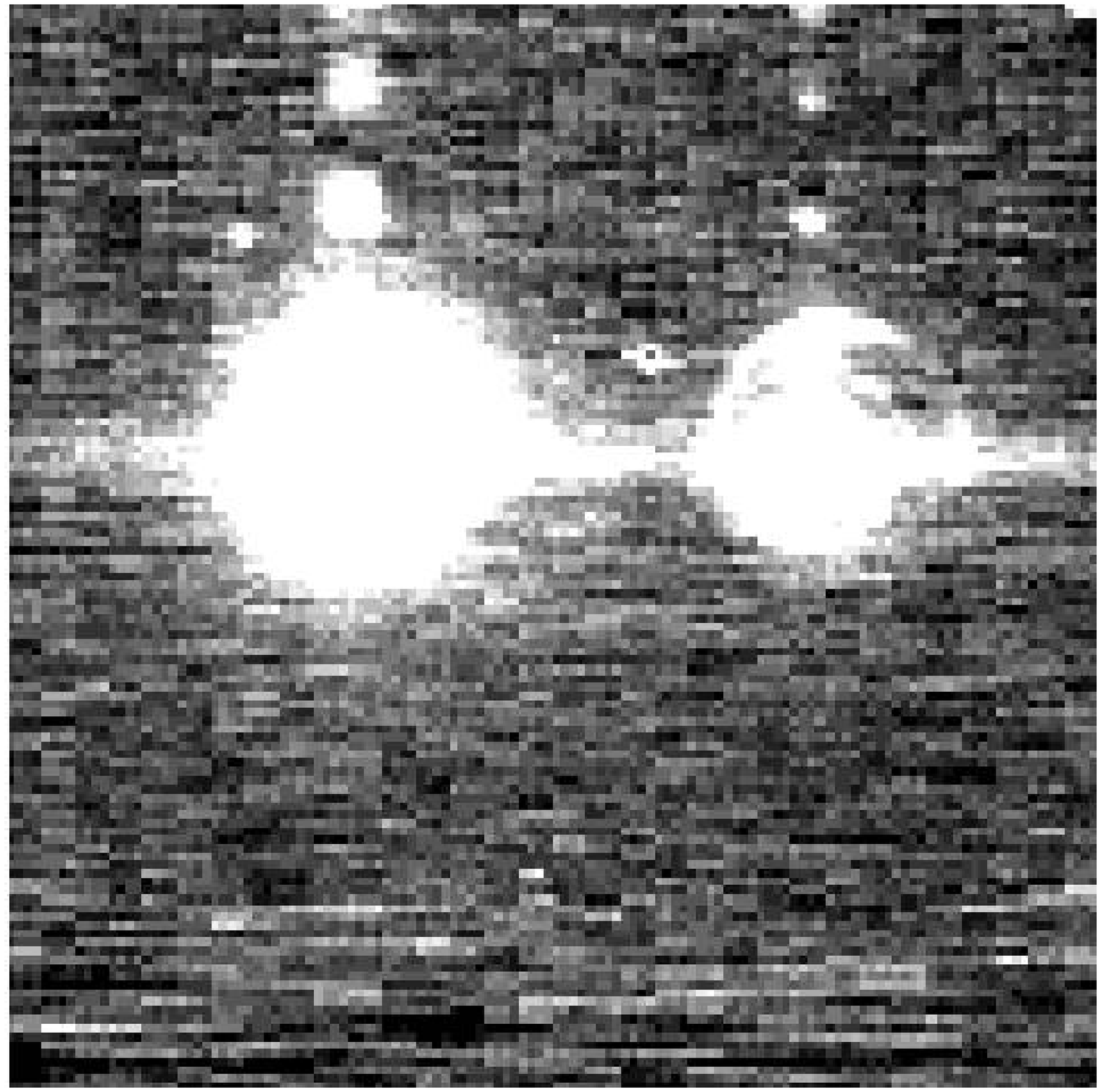}{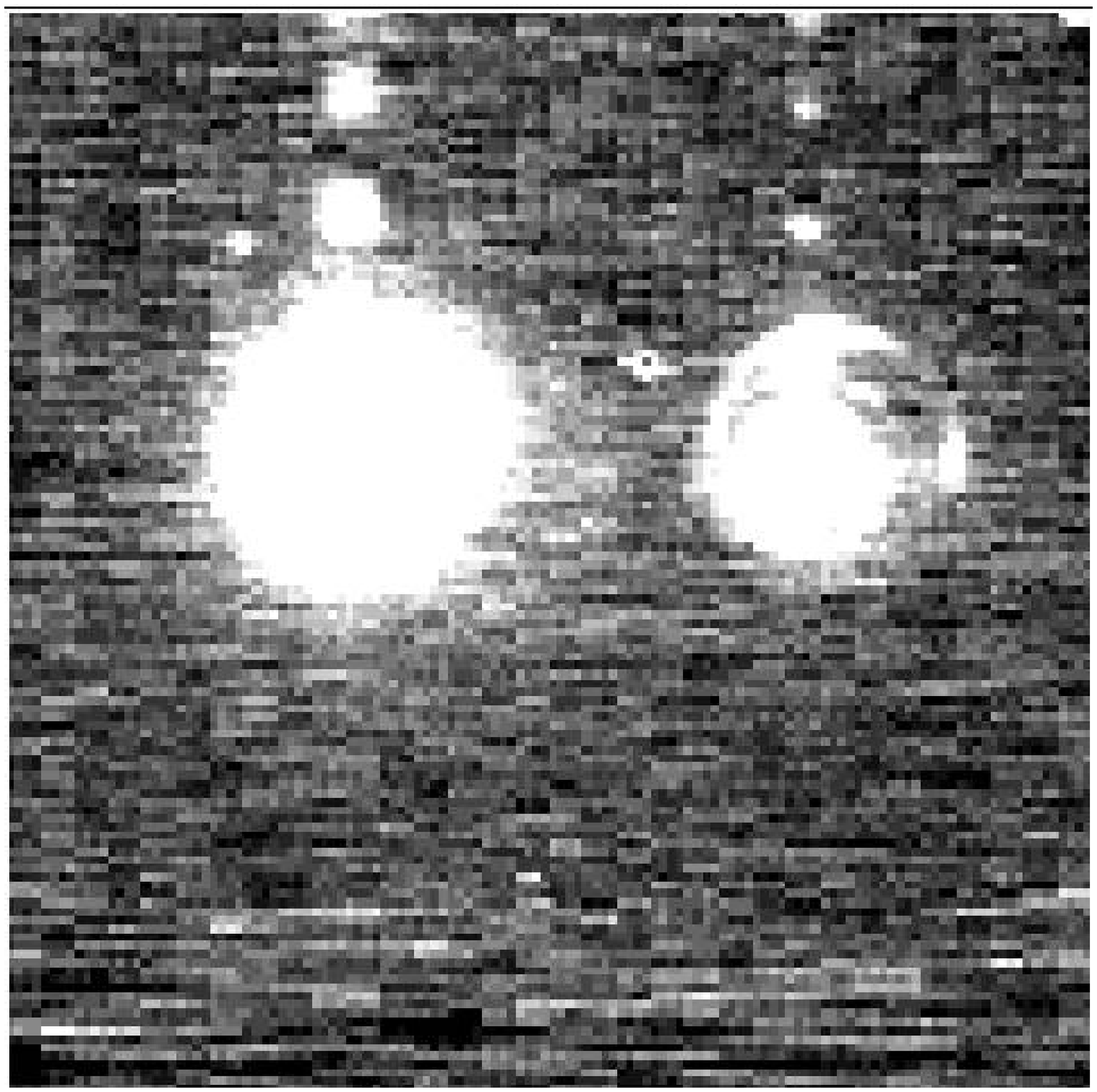}
\caption{Example of the rowdroop effect on the 24~\micron\ array.
Two pinhole sources were illuminated by an external stimulator source,
producing point sources on the array.  On the left panel, the rowdroop
signal produced by the pinhole sources can be seen as a horizontal
stripe of artificial signal (other structure around the point sources,
such as the arc around the right-hand source, is due to reflected
light from the pinhole apparatus, while the other apparent point
sources on the upper half of the array are latent signals produced by
earlier exposures of the pinhole sources at different positions).  The
right panel shows the same image, after the rowdroop correction has
been applied.
\label{fig_rowdroop}}
\end{figure*}

The rowdroop effect manifests itself as an additive constant to each
individual pixel and is proportional to the sum of the number of
counts measured by all pixels on its row, where a row is in the
cross-readout direction.  This effect is not completely understood,
and is similar to (but separate from) the droop phenomenon (see
\S\ref{sec_droop}).  The additive signal imparted to each pixel on a
row is constant and exhibits no gradient or dependence with pixel
position, thus, it is not related to a charge--bleed, or ``Mux--bleed''
effect.  The rowdroop contributes a small amount to the flux of an
individual pixel, and will only significantly affect pixels on rows
with high-intensity sources.  An example of rowdroop from ground-based
testing is shown in Fig.~\ref{fig_rowdroop}.

Using images of pinhole sources obtained in ground testing, we have
computed the row droop constant of proportionality, $K_\mathrm{rd}$.
This is the factor that gives the fraction of the total counts in a
row which is the result of row droop and should be subtracted from
each pixel in that row.  We find that the constant of proportionality for
the MIPS 24~\micron\ array is $K_\mathrm{rd} = 7.6 \pm 2.5 \times
10^{-5}$.  Thus, the rowdroop contributes $\approx 1$\% of the total
number of counts on a row.  The rowdroop is corrected for on a
read-by-read basis.

\subsubsection{Droop Subtraction}
\label{sec_droop}

Droop is a constant signal added to each pixel by the readouts.  The
exact cause of droop is unclear.  This extraneous signal, akin to a DC offset,
is directly proportional to the total number of counts on the entire
array at any given time.  We have measured the constant of
proportionality from ground test data.  The
droop coupling constant was measured to be $0.33 \pm 0.01$, which
agrees well with the 0.32 determined by BNA.

The droop correction algorithm first computes the mean signal on the
array, which is then multiplied by the droop coupling constant to
derive the droop signal, as given by
\begin{equation}
F_d = {{\sum_{ij} F_{ij}} \over {N_{pix}}} {{C_d} \over {1+C_d}} ,
\end{equation}
where $F_d$ is the droop signal, $F_{ij}$ is the signal on each pixel
(comprising both the actual incident flux and the droop), $N_{pix}$ is
the number of pixels, and $C_d$ is the droop coupling constant.  The
resultant droop signal is then subtracted from the original signal on
each pixel.

Under normal circumstances, the uncertainty associated with this
process is at the $\sim 1\%$ level, limited mainly by the uncertainty
on the coupling constant.  However, greater uncertainties arise when
pixels are saturated; since ADC saturation occurs well before hard
detector saturation, droop signal will still accumulate for an
incident flux above the ADC saturation level.  In this case, the
actual signal ramp must be extrapolated beyond the saturation point.
The droop signal is determined by extrapolating a fit to the
unsaturated portion of the ramp.  As with the rowdroop correction, the
droop correction is done on a read-by-read basis for RAW mode
24~\micron\ data.

\subsubsection{Dark Subtraction}
\label{sec_24dark}

Dark subtraction is done at each read using a dark calibration image
containing the full dark ramp for each pixel.  This step serves both
to remove the (small) dark current contribution and the
offset ramp starting points, so that each ramp starts near zero.

\subsection{Steps for Si SUR Mode Only}

The majority of the 24~\micron\ data are taken in the SUR mode
instead of the RAW mode.  In the SUR mode, a line is fit to the data
ramp on-board the spacecraft.  The resulting slope and first
difference (difference between the first two reads of the data ramp)
images are downlinked instead of the full ramp.  The first difference
frame effectively increases the dynamic range of the SUR mode as
signals that saturate somewhere in the ramp, but after the second
read, will have a valid measurement in the first difference frame.  To
reduce the data downlinked, any first difference value that is from a
ramp which does not saturate is set to zero.  This increases the
compressibility of the first difference frame.

\subsubsection{SUR Saturation Detection}

There can be degeneracy of SUR slope values due to the possibility of
saturation.  The possible slope value for a given pixel reaches a
maximum at full well, the point of ADC saturation.  After that point,
as the data ramp reaches saturation at the last few reads, the slope
value will begin to decrease because the on-board SUR algorithm does
not reject saturated reads.  In cases of extreme saturation, the slope
becomes quite small, and can eventually become zero if saturation
occurs within the first few reads.  The first difference value is
provided to break this degeneracy.  We have employed a conservative
threshold value for the first difference, above which a pixel is
flagged as being likely saturated.  ADC saturation occurs at +32768 DN
(see Figure~\ref{fig_24ramp} for an example of a saturated RAW ramp).
Assuming a linear ramp, the first difference for a ramp that just
saturates on the final read would be $65536/n_{read}$, where
$n_{read}$ is the total number of reads in the data ramp.  For
example, there are 60 reads in a 30 second DCE, yielding an ideal
saturation threshold of $\sim 1100$ DN/read.  To be more conservative,
we actually employ a threshold value of 1000 DN/read for a 30 second
exposure time and scale this for other exposure times.  Since the
data ramps are not linear, the actual first difference threshold is
larger than our chosen default value, so most cases of saturation will
be flagged.  The only exception being
saturation at the first read, in which case both the slope and the
first difference would be zero.  For all pixels that have been
flagged for saturation, the first difference value should be used in
place of the slope.

\subsubsection{Rowdroop, Droop, and Dark Subtraction}

The rowdroop and droop subtraction is done in the same way for SUR mode
as for RAW mode, except that the corrections are performed on the slope and
first difference images.

\subsubsection{Electronic Nonlinearity Correction}

Because the SUR data do not preserve the actual data ramps, the
linearity correction made somewhat complicated.  Nevertheless, the
quadratic behavior of the ramps can be used to analytically determine
the linearization of the SUR slope values.  This correction depends on
the observed SUR slope value, exposure time, and known quadratic
nonlinearity.  Note that saturation invalidates this method, as the
SUR slope-fitting algorithm does not reject saturated reads.  In this
case, no linearity correction is applied.

\section{Slope Image Calibration}
\label{sec_slope_cal}

The next step in the MIPS data reduction is to calibrate the slope
images while removing instrumental effects with time constants longer
than the DCE exposure times.  The instrumental effects corrected at
this stage include latents (24~\micron), responsivity drift (70 and
160~\micron), pixel-to-pixel responsivity variations, the telescope
illumination pattern, and flux nonlinearities (70 and 160~\micron).
The graphical representation of the data processing shown in
Fig.~\ref{fig_flowchart} gives the ordering of the reduction steps.
Since the time dependent responsivity of the Ge arrays requires
additional calibration steps than is usual for more common array
detectors, we give the mathematical basis of our Ge slope
calibration in \S\ref{sec_slope_math}.

\subsection{Principles}
\label{sec_slope_math}

Ignoring the 70 and 160~\micron\ flux nonlinearities, an uncalibrated
slope image can be represented by
\begin{equation}
\label{eq:SciFrame}
U(i,j,t_n) = [I(i,j)O(i,j) + D(i,j)] R(i,j,t_n)
\end{equation}
where $I(i,j)$ is the science image of interest, $O(i,j)$ represents
the telescope and instrument optics (the mean of $O(i,j)$ is one),
$D(i,j)$ is the dark current, and 
$R(i,j,t)$ is the instantaneous responsivity of the array; $i,j$
represent the pixel coordinates and $t_n$ the time of the $n$th DCE.
Calibration involves isolating $I(i,j)$, the flux from the
sky$+$object in the above equation.  The term $O(i,j)R(i,j,t_n)$ is
the equivalent of a traditional flat-field term.  As $R(i,j,t_n)$ is a
rather sensitive function of time for the 70 and 160~\micron\
detectors, however, a global ``flat-field'' cannot be determined, but
must be derived for each DCE separately.  The stimulators provide the
means to monitor $R(i,j,t_n)$ and all science observations will be
bracketed by stim flashes.  Stim flash images will be equivalent to
science frames with the addition of a stimulator illumination pattern:
\begin{equation}
\label{eq:stim}
U_{stim,N} = [S(i,j) + I(i,j)O(i,j) + D(i,j)] R(i,j,t_N)
\end{equation}
where $S(i,j)$ is the illumination pattern introduced on the array by the
stim flash with the mean of $S(i,j)$ equal to one.  MIPS observations
include the requirement that each 
stimulator flash will be preceded by a background exposure with the
identical telescope pointing; thus for the $N^{th}$ stimulator DCE
there exists a background DCE taken at time $t_{N} - \epsilon$,
\begin{equation}
U_{bkgd,N} = [I(i,j)O(i,j) + D(i,j)] R(i,j,t_N - \epsilon).
\end{equation}
If we assume that the responsivity of the array $R(i,j,t)$ doesn't
change dramatically between times $t_N$ and $t_N-\epsilon$,
i.e.\ $R(i,j,t_N) \sim R(i,j,t_N-\epsilon)$, we can construct for each
stimulator flash a background subtracted stim flash:
\begin{equation}
\label{eq:BkSubStim}
U_{stim,N} - U_{bkgd,N-\epsilon} \sim S(i,j) R(i,j,t_N).
\end{equation}
With background subtracted stim flashes determined from
Eq. \ref{eq:BkSubStim} for all stim flashes in the data set, an
instantaneous stim can be determined for any time, $t_n$, by
interpolation from bracketing stim flashes:
\begin{equation}
\label{eq:inst_stim}
S(i,j)R(i,j,t_n) = F[S(i,j)R(i,j,t_N)]
\end{equation}
where $F[]$ is some interpolating function on background subtracted
stims for times $t_N$ bracketing $t_n$.  Analysis of Ge
characterization array data indicates that a weighted linear fit
(weighted by the uncertainty in the stim flash frames) to two stim
flashes on either side of the data frame (a total of four stim
flashes) provides the optimal strategy for determining the
instantaneous stim amplitude (repeatability to $\sim$1\%\ on most
backgrounds).  Dividing science frames, Eq.  \ref{eq:SciFrame}, by the
interpolated instantaneous stim, Eq.  \ref{eq:inst_stim}, produces
\begin{equation}
\label{eq:ResCor}
U_{data}(i,j) = [I_{data}(i,j)O(i,j) + D(i,j)]/S(i,j).
\end{equation}
While we have removed the time dependent responsivity variation, the
data of interest, $I_{data}(i,j)$, are still modified by the optical
response and the dark current; in addition, we have introduced the
stimulator illumination pattern into our data.  Fortunately, since the
time dependence has been removed, we can remove these other
instrumental signatures through carefully accumulated calibration
data.

First, the dark correction, $D(i,j)$, can be determined from a
sequence of exposures as above, with the additional constraint that
the scan mirror be positioned such that no light from the ``sky''
falls on the detector. Thus the data and stim flashes in a dark
current data sequence are represented by
\begin{equation}
\label{eq:dark}
U(i,j,t_n) = D(i,j)R(i,j,t_n)
\end{equation}
and
\begin{equation}
\label{eq:dark_stim}
U_{stim} = [S(i,j) + D(i,j)] R(i,j,t_N),
\end{equation}
respectively.  The dark data are corrected for responsivity variations
exactly as described above and the individual frames combined to
produce an average dark current, $D(i,j)/S(i,j)$. Subtracting this
dark current from science frames that have been corrected for
responsivity variations, Eq. \ref{eq:ResCor} yields
\begin{equation}
\label{eq:DarkResCor}
U_{data}(i,j) = \frac{I_{data}(i,j)O(i,j)}{S(i,j)},
\end{equation}
our responsivity, dark corrected science frame.  What remains is to
correct for the telescope optics, $O(i,j)$, and the stim illumination
pattern, $S(i,j)$.

Correcting for the combined illumination pattern of the telescope and
stim involves a standard series of MIPS exposures, i.e. data frames
interspersed with stim flashes.  As such, they may be represented by
equations of the form Eq. \ref{eq:SciFrame}, where the $I_n(i,j)$
represent dithered images of ``blank'' sky fields.  Calibrating the
sequence by correcting for responsivity variations and dark current as
above results in a series of images
\begin{equation}
U_{illum}(i,j,t_n) = \frac{I_n(i,j)O(i,j)}{S(i,j)}.
\end{equation}
Since by construction, the $I_n(i,j)$ are dithered images of
``smooth'' regions, if a large number of $I_n(i,j)$ are acquired, they
may be median combined to remove point sources (and extended sources
if dithered ``sufficiently''), cosmic rays, etc.  resulting in
\begin{equation}
\label{eq:IllCorPre}
U_{illum}(i,j) = \langle I_n(i,j) \rangle \frac{O(i,j)}{S(i,j)} = C
\frac{O(i,j)}{S(i,j)}
\end{equation}
where $O(i,j)$ and $S(i,j)$ are constant regardless of telescope
pointing.  Hence the median only affects the changing sky image as the
telescope is dithered.  The constant $C$ in equation \ref{eq:IllCor}
may be set to one resulting in the illumination correction frame
\begin{equation}
\label{eq:IllCor}
U_{illum}(i,j) = \frac{O(i,j)}{S(i,j)}.
\end{equation}
The responsivity corrected, dark subtracted data (equation
\ref{eq:DarkResCor}) are now divided by the illumination correction
resulting in
\begin{equation}
\label{eq:CalibratedData}
U_{data}(i,j) = \frac{I(i,j)O(i,j)}{S(i,j)} / \frac{O(i,j)}{S(i,j)} =
I(i,j),
\end{equation}
and we have recovered the quantity of interest, the astronomical sky
$I(i,j)$.  Suitable observations of standard stars can then be used to
convert instrumental counts to physical units (e.g. Janskies).

\subsection{Dark, Flat Field, and Illumination Correction}

\begin{figure*}
\plotone{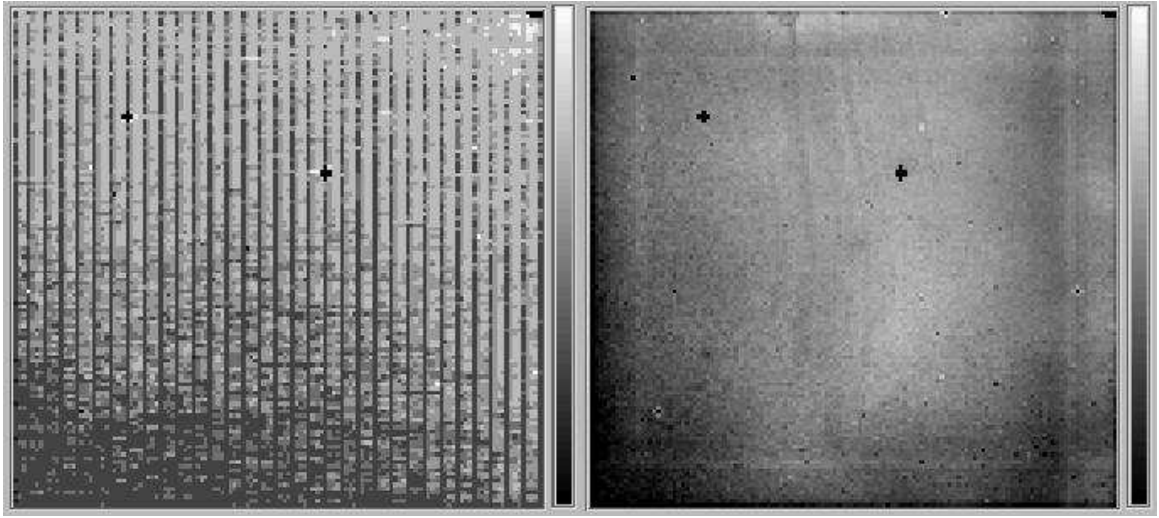}
\caption{Calibration Images for the 24~\micron\ array.  They
are the dark (left) and flat field (right).  The displayed range for
the dark image is 0 to 4 DN/s.  The displayed range for the flat field
image is 0.9 to 1.1.  The pixels displayed as solid black correspond
to the handful of known bad pixels.  \label{fig_24cal}}
\end{figure*}

\begin{figure*}
\plotone{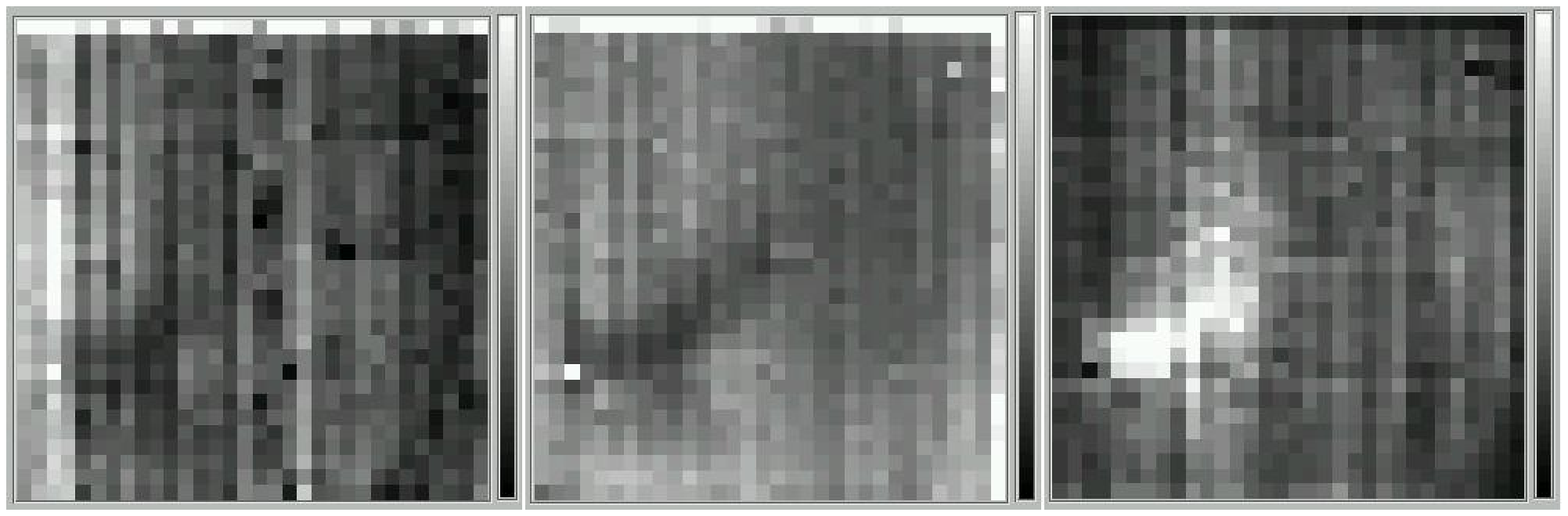}
\caption{Calibration Images for the 70~\micron\ array.  They
are the dark (left), the illumination correction (middle), and an
example of a stim flash (right).  The displayed range for the dark
image is 0 to 0.02.  The displayed range for the illumination
correction image is 0 to 2.0.  The displayed range for the stim flash
is 0 to 40,000 DN. \label{fig_70cal}}
\end{figure*}

\begin{figure}
\plotone{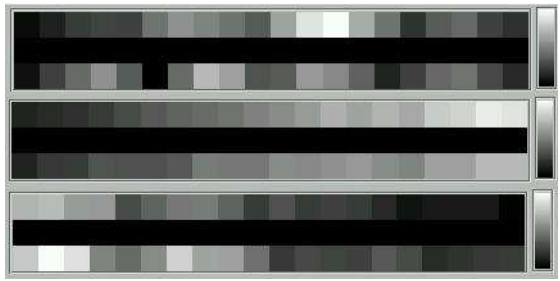}
\caption{Calibration Images for the 160~\micron\ array.  They
are the dark (top), the illumination correction (middle), and an
example of a stim flash (bottom).  The displayed range for the dark
image is 0 to 0.65.  The displayed range for the illumination
correction image is 0 to 2.0.  The displayed range for the stim flash
is 4,000 to 24,000 DN. \label{fig_160cal}}
\end{figure}

The dark, flat field, and illumination correction calibration images
described above will be obtained throughout the life of the mission.
Example preflight calibration images are shown in
Figs.~\ref{fig_24cal}-\ref{fig_160cal}.  Simulations indicate that
high S/N flat field and illumination correction images ($\sim 0.5\%$
RMS) can be obtained with dithered observations of ``smooth'' areas of
the sky: $\sim$60--100~DCEs at 24~\micron\ and $\sim$200~DCEs at
70~\micron\ are required.  At 160~\micron, the situation is less
ideal, with simulations indicating as many as 500~DCEs may be required
to produce flats to better than 1\%\ RMS.

\subsection{Si Latent Correction}

Si IBC arrays are known to have considerable latency, where the signal
induced by bright illumination persists after the illumination has
terminated.  Ideally, if one knows the position of a source exposed on
the array and the latency decay behavior, these artifacts can be
subtracted from an image.  We have characterized the latent behavior
from ground test data.  Several different conditions were
explored, including varying brightnesses of the illuminating source,
varying brightnesses of the background, initial
bias boosts, and changing the number of resets via different exposure
times.  A bias boost can flush out most of the trapped charge, but
resets are not nearly as effective.  Since bias boosts will only be
done in the first DCE of each observation, we correct for
latent residuals in the data processing.

The latent decay curve can be described by single exponential, given
by
\begin{equation}
m(t) = m_o + p e ^{-t/\tau} ,
\end{equation}
where $m_o$ is the slope in the absence of a latent, $p$ is the
initial value of the latent, and $\tau$ is the latent time constant.
Based on the limited ground data, the latent parameters ($p$ and
$\tau$) appear to be functions of background levels, number of resets and
possibly location on the array. In general, the latent contribution is
about $\sim$ 1\% of the initial source brightness $\sim$5 sec after
that source has shut off.  Higher background yield slightly higher
values for $p$ and lower values of $\tau$.  The value of $\tau$ is in
the range of $12 \pm 5$ seconds. 

\subsection{Ge Flux Nonlinearity Correction}

Both the 70 and 160~\micron\ arrays exhibit nonlinearities that are
dependent on the incident point source flux as well as the background.
These are termed flux nonlinearities and have been
observed in data taken with the characterization array as well as the
flight array.  As is usual for the Ge arrays, each pixel shows flux
nonlinearities with a different dependence on source flux and
background.  Correcting for this
effect can be broken into two pieces: (1) removing the pixel to pixel
differences in the nonlinearity followed by (2) the application of a
global nonlinearity correction as a function of the source brightness
and background.  

The pixel to pixel variations in the flux
non-linearity may be mapped by analyzing the ratio of two stim
flashes, where one is the standard on orbit calibrating stim flash.
Measured differences in the ratio from pixel to pixel can be used to
correct each pixel to the same flux non linearity for the given
background and source (second stim flash) amplitude. Repeating the
measurement for a variety of second stim flash amplitudes (up to
saturation for each pixel) and backgrounds will map out the
correction.  The second, global, stage of the correction can be
characterized by observations of calibration stars with a range of
known brightness ratios on similar backgrounds.  The combination of
these two tasks outlined above should provide a good measurement of the
flux nonlinearity correction for a range of backgrounds.  This
correction will improve continuously during the mission as the range
of backgrounds and calibration stars expands.

\subsection{Flux Calibration}

The absolute calibration of MIPS will rely on a well determined anchor
at 10.6~\micron\ using the fundamental calibrators $\alpha$~Boo,
$\alpha$~Tau, and $\beta$~Gem \citep{rieke85,cohen92}.  Three
independent methods will be used to extrapolate the calibration at
10.6\micron\ to the MIPS bands: (1) Solar analogs, (2) A star
atmospheric models, and (3) semi-empirical models of K giants.  Grids
of stars for each method have been observed from the ground and tied
to the fundamental calibrators at 10.6~\micron.  For the solar analog
stars, on orbit observations at 24, 70, and 160~\micron\ are being
compared with extrapolations of empirical measurements of the sun
extrapolated into the MIPS bands. A grid of A stars has been observed
in all three bands on orbit and compared to extrapolations of A star
atmosphere models to the MIPS bands.  While the solar analog and A
star calibrators will be observed in the MIPS 160~\micron\ band, the K
giant calibrators will be the only ones detectable at high
signal-to-noise in that band. On orbit observations of the K giant
calibrators are being compared to theoretical extrapolations of model
atmospheres, eg. \citet{cohen95, cohen96a, cohen96b} extrapolated to
longer wavelengths using the Engelke function \citep{engelke92}.
Absolute flux calibrators will be observed throughout the lifetime of the
mission.

\section{Using Redundancy to Improve Calibration}
\label{sec_redund}

The last step in the reduction of MIPS data is to use the redundancy
inherent in the observing modes to improve the removal of instrumental
signatures.  This step is mainly for the 70 and 160~\micron\ data due
to the challenging aspects of Ge detector calibration.  We define the
level of redundancy to be the number of different pixels that measure
the same point on the sky.  Our approach will be to look for known instrumental
signatures (as a function of time) in the difference between what a
particular pixel detects and what all the other pixels detected for
the same sky locations.  This is possible because the observing
strategy has been designed so that each point on the sky will be
observed multiple times by different pixels.

Table~\ref{tab_redund} shows the minimum level of redundancy for each
MIPS observing mode.  Many MIPS observations are taken with
multiple cycles resulting in significantly higher redundancies.  It is
recommended to have a minimum redundancy of four.

\begin{deluxetable*}{lccc}
\tablewidth{0pt}
\tablecaption{Redundancy in MIPS Observing Modes\label{tab_redund}}
\tablehead{ \colhead{Type of AOR} & \colhead{24~\micron} & 
   \colhead{70~\micron} & \colhead{160~\micron} }
\startdata
Photometry, compact                   & 14         & 10       & 2 \\ 
Photometry, large                     & 10         & 6        & 1 \\  
Photometry, compact super resolution  & 14         & 8        & 6  \\       
Photometry, large super resolution    & 10         & 8        & \nodata \\

Scan Map, slow                        & 10         & 10       & 1 \\
Scan Map, medium                      & 10         & 10       & 1 \\
Scan Map, fast                        & 5          & 5        & 0.5 \\

SED                                   & \nodata    & 2        & \nodata \\
Total Power                           & 1          & 1        & 1 
\enddata
\end{deluxetable*} 

\subsection{Algorithm}
 \label{sec_redund_algorithm}

The basic algorithm for using redundancy to refine the instrumental
signature removal is as follows.

\begin{enumerate}
\item Create a mosaic of all the images in question.  During the
  mosaic creation, use a sigma rejection algorithm to remove data
  that are deviant from the majority of the observations.
\item Use the mosaic as a ``truth'' image of what each image should have
  measured.
\item For each pixel, subtract the actual from the ``truth''
  measurements to create a measurement of the time history
  differences.
\item Examine the difference time history for known instrumental
  signatures.  While many instrumental signatures could be present, we
  plan to concentrate on stim latent residuals and systematic
  differences between ``extended'' sources and point sources.  Actual
  on-orbit data will guide the details and number of instrumental
  signatures that are corrected using redundancy.
\item Correct for all instrumental signatures that are found to be
  significant.
\item Iterate steps 1-5 until no new significant instrumental
  signatures are found.
\end{enumerate}

The input to this algorithm is calibrated slope images.  The output
product of this algorithm is enhanced images.  A useful side product
will be the mosaicked image of the object.

\subsection{Distortions of Arrays}

To use the redundancy to remove additional instrumental signatures we
must first coadd all related observations into a single mosaic.
Because the MIPS optical train is made up of purely off-axis
reflective elements there exist scale changes and rotations across the
re-imaged focal plane. To coadd images taken at different places on
the array, it is crucial to correct the data for these distortions.

\begin{figure}
\plotone{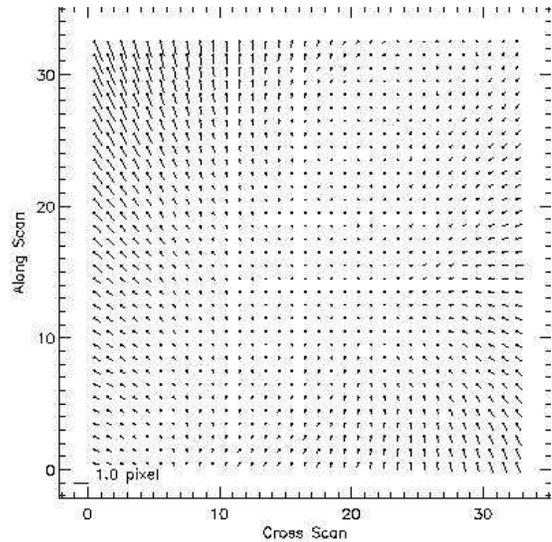}
\caption{70 $\mu$m narrow field mode residuals between the distorted
and undistorted points after the object angles were converted to
pixels.  Note the 1 pixel scale in the lower left-hand
corner. \label{fig_70um_dist}}
\end{figure}

We used the Code V optical models for Spitzer/MIPS to estimate the
distortions present in the images from the three MIPS detectors.  The
results from Code V allow us to determine distortion polynomials
which can then be used to correct for the distortions.  We estimated
the distortions by setting up a grid of equally 
spaced points in the field of view at a specific scan mirror
angle. The chief ray from each object point was traced through the
system to where it was imaged on the focal plane. In a perfect optical
system the image points would map perfectly from the object with a
possible change in magnification. The difference between the ideal
location and the actual location is the distortion.  For example,
Figure~\ref{fig_70um_dist} is a vector plot of the distortions present
in the 70~\micron\ narrow field array. The equally spaced grid of
points present the focal plane points and the ends of the vectors
correspond to the object points, after a plate scale factor was
applied. The difference in the points (the length of the vector) is
caused by the distortions.

\begin{deluxetable}{lc}
\tablewidth{0pt}
\tablecaption{Distortions in MIPS Detectors \label{tab_dist_det}}
\tablehead{ \colhead{Detector} & \colhead{\% FOV Scale} }
\startdata
 24 \micron\    & 2.84  \\
 70 \micron\ Wide & 0.2   \\
 70 \micron\ Narrow & 7.70  \\
 160 \micron\   & 7.78  \\
\enddata
\end{deluxetable}

\begin{figure}
\plotone{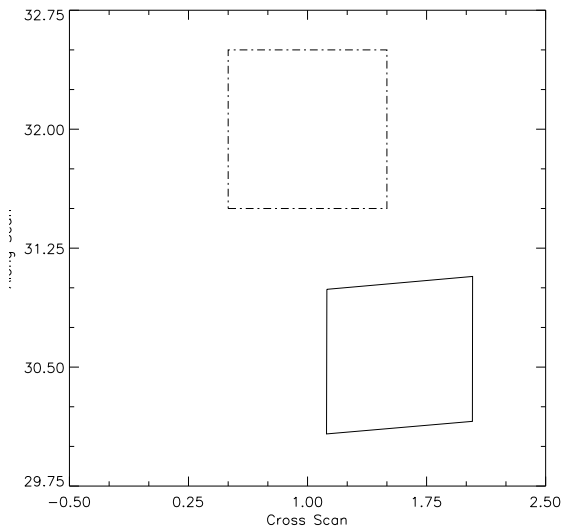}
\caption {70 $\mu$m NF mode distorted pixel. This pixel is located
in the right hand corner of the array with (area distorted/area undistorted) =
1.1929.  In this plot the distorted pixel is plotted with a solid line
and the undistorted pixel is plotted with a dot-dash
line. \label{fig_70um_close}}
\end{figure}

Table~\ref{tab_dist_det} lists the scale change of the field of view
of the different MIPS arrays. The scale change is defined as (maximum
length of distorted field - minimum length of the distorted
field)/(minimum length of the distorted field).  From a distortion
standpoint, it is useful to look closely at individual pixels to see
how distortion changes the area imaged on the pixel.
Figure~\ref{fig_70um_close} is a plot of a distorted pixel in the
70~\micron\ narrow field array. One can see that the distorted pixel changes
shape from a square to a somewhat trapezoidal shape. The ratio of the
distorted to undistorted pixel area is 1.19.  Table~\ref{tab_dist_pix}
lists information on how distortion affects the area imaged on
individual pixels on the different arrays.  The distorted pixel area
ratio is defined as (distorted pixel area)/(undistorted pixel area).

\begin{deluxetable*}{lccccc}
\tablewidth{0pt}
\tablecaption{Distorted Pixel Area Ratio \label{tab_dist_pix}}
\tablehead{ & \multicolumn{4}{c}{Area Ratio} \\
  \colhead{Detector} & \colhead{Mean} &
  \colhead{Std. Dev.} & \colhead{Min.} & \colhead{Max.} }
\startdata
 24 \micron\          & 0.9998  & 0.0282  & 0.9406  & 1.0613 \\
 70 \micron\ Wide     & 1.0027 & 0.0042  & 0.9973  & 1.0148 \\
 70 \micron\ Narrow   & 1.0129 & 0.0664  & 0.8913  & 1.1929 \\
 160 \micron\         & 0.9781  & 0.0361  & 0.9007  & 1.0137 \\
\enddata
\end{deluxetable*}

Following the procedure of converting the pixel coordinates to world
coordinates outlined in \citet{gre02}, the distortion correction is
applied to the pixel coordinates before any other transformations.
The distortion correction is accounted for by distortion polynomials.
The distortion polynomials give the additive correction to map the
distorted pixel coordinates, $u,v$ to the distortion corrected pixel
coordinates $p,q$.  Thus, $p = u + F(u,v) $ and $q = v + G(u,v)$,
where
\begin{equation}
F(u,v) = A_{20}u^2 + A_{02}v^2 + A_{11}uv + A_{30}u^3 + \\ A_{21}u^2v +
A_{12}vu^2 + A_{03}v^3
\end{equation}
and
\begin{equation}
G(u,v) = B_{20}u^2 + B_{02}v^2 + B_{11}uv + B_{30}u^3 + \\ B_{21}u^2v +
B_{12}vu^2 + B_{03}v^3 .
\end{equation}

\subsection{Mosaicking Details}

\begin{figure}[tbp]
\plotone{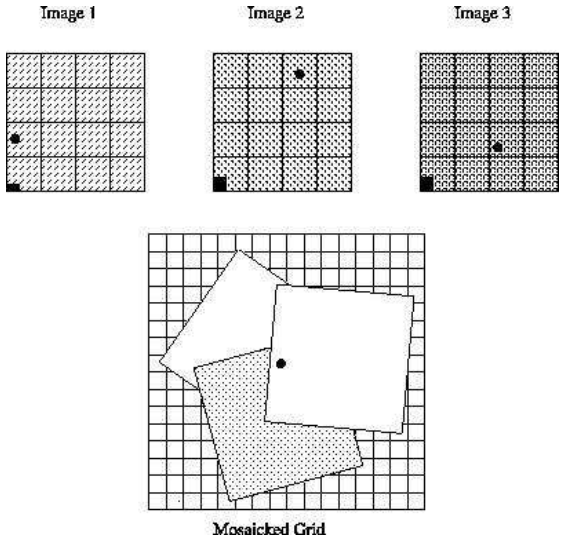}
\caption{An example of how a single pixel from three different images
(only $4 \times 4$ pixels shown) are overlapped on the mosaic image is
shown.  The solid point represents the same location on the sky as it
would be observed in each image.  \label{fig_mosaic}}
\end{figure}

The ability to remove additional instrumental signatures is dependent
on creating a high resolution mosaicked image. For the mosaicked image
to be of sufficient resolution, the mosaicked pixel sizes must be
smaller than the original input pixels.  While many mosaicking programs compensate for
undersampling by making use of dithered data, the MIPS data are well sampled
and do not require this compensation.  Instead, our focus is co-adding the related
calibrated images into a single image without interpolating between
pixels. Therefore, we always work on the coordinates of the corners of
a pixel, transforming them from the input image coordinate
system to the output mosaicked coordinate system. The output mosaic
image is on a single tangent plane.  In the transformation of the
pixel corners in the input image pixels to their location on the
output mosaic image the corners are corrected for distortion,
converted to right ascension and declination, and then projected onto
the tangent plane defined by the right ascension and declination of the
mosaic center.  Figure~\ref{fig_mosaic} is an example of three images
which overlap each other on the mosaicked plane.  In the process of
establishing the location of the image pixel corners on the mosaicked
plane, the link and overlap coverage between the input pixel and the
output pixels it falls on is determined. A critical step in removing
residual instrumental signatures based on the co-added mosaic image
depends on correctly linking each mosaic pixel with each image pixel
that overlaps it (and vice versa) and accurately determining the
degree of overlap. Essentially each output sub-sampled mosaic pixel
becomes a cube of data, with each plane in this cube representing the
information in each overlapping image pixel. The surface brightness
and uncertainty associated with each mosaic pixel is found by weighted
averaging the overlapping planes of data.  In the surface brightness
case, the weighting is based on the overlap coverage and uncertainty
associated with the input image pixel.


The information in each mosaic pixel is based on multiple observations
of a single area on the sky. This redundancy of data can be used to
identify cosmic rays or any single image pixel measurement that
deviates from the expected mean of multiple observations and expected
noise.  For example, for a 70~\micron\ photometry observing mode
cycle, if one of the pixels suffers from a much larger stim latent
than the other observations, it will stand out and be identified as an
outlier. As an outlier, it will not be used in creating the mosaicked
image. After all the outliers have been determined, then the links
between the output mosaic pixel and the input image pixels are used to
tally the number of times an image pixel was flagged as an outlier. If
the majority of the time an image pixel was flagged as deviant, then
this pixel is flagged in the original data as an outlier. If a
sufficiently large number (about 1\%) of the input image pixels are
flagged as outliers then the mosaic step is repeated. The final output
is a mosaic image that can then be used as the ``truth'' image of
what each image should have measured.  Following the steps outlined in
section~\ref{sec_redund_algorithm} this truth image is used to remove
residual instrumental signatures.

\section{Initial Testing with Flight Data}
\label{sec_inflight}

With the successful launch of the Spitzer Space Telescope in August
2003, these reduction algorithms were tested against MIPS flight data
of astronomical sources.  This testing has validated the
algorithms described in this paper, but has also shown that a number of
modifications will be needed to handle the realities of flight data.
The initial results of this testing are summarized here and in
\citet{gor04}, but a full accounting will be a subject of a future
paper when the final MIPS reduction algorithms are known.

There were two significant changes in the instrument operations which
are not easily correctable by reduction algorithms.  The 70~$\micron$
array was found to suffer from a cable short induced sometime between
ground testing and flight.  This short injects a large amount of noise
into one half of the 70~$\micron$ array resulting in a useful
array of only $16 \times 32$~pixels.  The 160~$\micron$ array was
found to suffer from a ``blue-leak'' caused by an unintended
reflection from the blocking filter which passes through the bandpass
filter.  This ``blue-leak'' results in an approximately factor of 15
image leak for stellar sources.  This leak means that asteroids are
now the primary calibrators for 160~$\micron$.  Other than bright
stars, the leak signal is below the confusion limit for most science
targets as they have much smaller blue/160~$\micron$ ratios.

At 24~$\micron$, the pre-flight reduction algorithms were found to
work well with only three changes needed.  First, the Row Droop
Correction does not seem necessary, but extensive testing has yet to
be completed.  Second, an additive offset in the second read of every
ramp was found which produced a low level (1-2\%)
gradient in final mosaics.  A straightforward correction for this has
been implemented using RAW and SUR data for calibration.  Third, scan
mirror angle dependent flat fields are needed due to contamination of
the scan mirror by small particles.  This contamination is seen as
dark spots in individual images which move with scan mirror angle, but
not spacecraft offsets.  With these three modifications to the
preflight algorithms, MIPS is producing high quality 24~$\micron$
images which are well calibrated.

At 70~$\micron$, flight data has validated the basic structure
of the preflight reduction algorithms but significant modification is
required to account for time dependent behaviors.  The stim flash
latents were found to grow in amplitude quickly after anneals.  With a
similar timescale, the residual background time dependence (after
correction using the stim flash amplitudes) was seen to grow.  These
two facts required hand reductions to remove the stim flash latents and
background variations to produce good quality mosaics at
70~$\micron$.  These two effects are prime candidates for removal
using redundancy, but the effectiveness of automatic removal has not
been demonstrated yet.

At 160~$\micron$, the basic preflight algorithms have been validated
from comparison with flight data.  Some differences in detector
behavior were seen in flight data.  For example, the stim flash
latents have a faster time constant than in preflight data.  At this
time, the nonlinearities in the 70 and 160~$\micron$ arrays have not
been well enough characterized with flight data to validate this
section of the preflight algorithms.  Finally, the cosmic ray rate
seen in the Ge arrays has been seen to be about a factor of two over
preflight predictions, 1 cosmic ray every 12 or so seconds.  The ramp
jump detection has been seen to work well and line segment fitting
removes the majority of the effects of these cosmic rays.  Some
residual effects remain and additional characterization may lead to
algorithms to remove these additional effects.

\begin{figure*}
\plotone{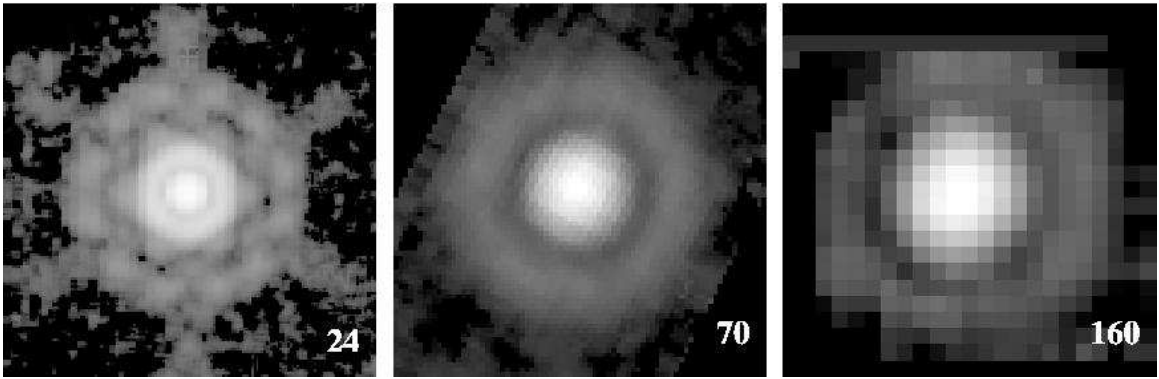}
\caption{Images of the 24, 70, and 160~\micron\ observed PSFs are
shown with dimensions of $60\arcsec \times 60\arcsec$, $100\arcsec
\times 100\arcsec$, $150\arcsec \times 150\arcsec$, respectively.  for
The 24~\micron\ PSF was created from observations of 
HD~159330 a K2III star with a predicted flux of 0.570~Jy.  The
70~\micron\ PSF was created from fine-scale observations of HD~131873
a K4III with a predicted flux of 3.14~Jy.  The 160~\micron\ PSF was
created from observations of Harmonia an asteroid with a predicted flux
of $\sim$1.5~Jy.
\label{fig_psf}} 
\end{figure*}

The effectiveness of the algorithms described in this paper as well as
the design of MIPS is attested by the point-spread-functions (PSFs)
constructed from flight data at 24, 70, and 160~$\micron$ shown in
Fig.~\ref{fig_psf}.  These PSFs all clearly have a well-defined first
Airy ring with the 24~\micron\ PSF also exhibiting a well-defined
second Airy ring.  All three PSFs are well represented by the
predictions of TinyTim models \citep{kri93} adapted to MIPS.  In
addition, there are many papers written using MIPS flight data for the
Spitzer Special Astrophysical Journal Supplement Issue (2004, ApJS,
154).




\section{Summary}
\label{sec_summary}

This paper has described the preflight data reduction algorithms for
all three arrays for the MIPS instrument on Spitzer.  These algorithms
have been guided by extensive laboratory testing of the Si
(24~\micron) and Ge (70 and 160~\micron) arrays.  In addition,
numerical modeling of the Ge arrays has provided important insights
into their behavior.

The design and operation of the MIPS instrument has been summarized
to give sufficient background for understanding the data reduction
algorithms.  The design and operation of the MIPS instrument is mainly
driven by the needs of the Ge arrays.  As Ge detectors display
significant responsivity drift over time due mainly to cosmic ray
damage, the MIPS observing modes include frequent observations of an
internal illumination source.  In addition, most MIPS operating modes
have been designed to provide significant redundancy to increase the
robustness of the MIPS observations against detector effects.

The data reduction for the MIPS arrays is divided into three parts.
The first part converts the data ramps into slope measurements and
removes detector signatures with time constants less than
approximately 10 seconds.  These detector signatures at 24~\micron\
include saturation, dark current, rowdroop, droop, electronic
nonlinearities, and cosmic rays.  At 70 and 160~\micron, the detector
signatures removed include saturation, electronic nonlinearities, stim
flash latents, and cosmic rays.  The resulting slopes are determined
from linear fits and their uncertainties are computed accounting for
both the random and correlated nature of the data ramp uncertainties.

The second part of the MIPS data reduction converts the slopes to
calibrated slopes and removes detector signatures with time constants
larger than approximately 10 seconds.  At 24~\micron, this translates
to applying a flat field, correcting for object latents, and applying
the flux calibration.  At 70 and 160~\micron, this step includes
subtracting the dark, flat fielding using an instantaneous flat field,
correcting for the flux nonlinearities, and applying the flux
calibration.  A flat field specific to each 70 and 160~\micron\ image
is required to correct for the time-dependent responsivity of the Ge
arrays.  It is constructed from the frequent stim flashes and a
previously determined illumination correction.

The third data reduction step is to use the spatial redundancy inherit
in the MIPS observing modes to improve the removal of instrumental
signatures.  This step is only applied to the Ge data.  Known
instrumental signatures are searched for in the difference between
what a specific pixel and what all other pixels from the same sky
locations detected.  If instrument signatures are detected, they are
removed and the process is repeated.  This method is iterative in
nature and will require care to avoid introducing spurious signals
into the data.  The design of this portion of the data reduction
algorithms is necessarily the least developed because only after
Spitzer launches will it be known which instrumental signatures are
important to correct with this method.

Finally, initial testing using flight data from MIPS has validated
these data reduction algorithms, but some modification is necessary to
account for the realities of flight.  A future paper will describe
these modifications in detail once they have been devised and tested.

\acknowledgements
We wish to thank J.\ W.\ Beeman and E.\ E.\ Haller for their
contributions to the design and building of the MIPS instrument.  This
work was supported by NASA JPL contract 960785.

\appendix

\section{Linear Fit to Data with Correlated and Random Uncertainties}
\label{app_fit_unc}

When a detector is non-destructively read out multiple 
times before resetting the resulting data ramps represent 
correlated measurements.  This is because measurement $y_{i+1}$ is
equal to $y_i 
+ p_i$ where $p_i$ is number of photons detected in the time between
$y_i$ and $y_{i+1}$.  This statement ignores the effects of read noise,
which produces uncorrelated uncertainties on the $y_i$ measurements.
While fitting lines to data with correlations is a complex subject,
the form of the correlations in the case of non-destructive readouts
allows analytic equations to be derived for the linear fit
parameters and uncertainties.  We present a derivation of equations
for linear fit parameters and their uncertainties for the case of a
data ramp with correlated reads and no read noise.  This derivation is
based on a similar derivation by \citet{spa98} for NICMOS data ramps
but is slightly more general.  As part of this derivation, it can be
seen that the linear fit parameters derived assuming either random or
correlated uncertainties are equivalent.  This is not the case for the
uncertainties on the fit parameters, which is the main motivation for
this derivation.

The basics of fitting a line to data with random uncertainties are
given in \citet{bev92}.  We repeat their results here, as the
derivation for correlated uncertainties draws directly from this work.
In fitting data to a line of the form
\begin{equation}
y_i = a + b x_i \label{eq_line}
\end{equation}
the fit parameters and their uncertainties are
\begin{eqnarray}
a & = & \frac{S_{xx} S_y - S_x S_{xy}}{\Delta}, \label{eq_a_random} \\
b & = & \frac{S S_{xy} - S_x S_y}{\Delta}, \label{eq_b_random} \\
\sigma_a(ran)^2 & = & \frac{S_{xx}}{\Delta},~~\mathrm{and} \\
\sigma_b(ran)^2 & = & \frac{S}{\Delta} \label{eq_sigma_b_random}
\end{eqnarray}
where $N$ is the number of ($x_i,y_i$) measurements, $\sigma(y_i)$ is
the uncertainty on each measurement of $y_i$, 
\begin{eqnarray}
S & = & \sum_{i=1}^N \frac{1}{\sigma(y_i)^2}, \\ 
S_x & = & \sum_{i=1}^N \frac{x_i}{\sigma(y_i)^2}, \\ 
S_{xx} & = & \sum_{i=1}^N \frac{x_i^2}{\sigma(y_i)^2}, \\ 
S_y & = & \sum_{i=1}^N \frac{y_i}{\sigma(y_i)^2}, \\ 
S_{xy} & = & \sum_{i=1}^N \frac{x_i y_i}{\sigma(y_i)^2},~~\mathrm{and} \\
\Delta & = & S S_{xx} - \left( S_x \right)^2. \label{eq_delta}
\end{eqnarray}
These equations assume that the measurements of $y_i$ are independent.

To determine the linear fit terms for a line fit to correlated
data ramps, the assumption that the $y_i$ measurements are independent
is not correct.  The standard formulae
need to be modified to sum over terms that are independent.  The
modifications start with realizing that 
\begin{equation}
p_i = y_{i} - y_{i-1}
\end{equation}
is the independent quantity in the absence of read noise.  Any
equation in the standard derivation that  
relies on the independence of $y_i$ needs to be modified to only
depend on $p_i$.  Thus,
\begin{eqnarray}
S_y & = & \sum_{i=1}^N \frac{y_i}{\sigma(y_i)^2} \\ 
  & = & \frac{y_1}{\sigma(y_1)^2} + \frac{y_1 + p_2}{\sigma(y_2)^2} +
     \frac{y_1 + p_2 + p_3}{\sigma(y_3)^2} + \cdots \\ 
  & = & y_1 \sum_{i=1}^N \frac{1}{\sigma(y_i)^2} + p_2 \sum_{i=2}^N
     \frac{1}{\sigma(y_i)^2} + p_3 \sum_{i=3}^N \frac{1}{\sigma(y_i)^2} +
     \cdots \\ 
  & = & y_1 \sum_{i=1}^N \frac{1}{\sigma(y_i)^2} +
     \sum_{i=2}^N \left( p_i \sum_{k=i}^N \frac{1}{\sigma(y_k)^2}
     \right) \\ 
  & = & y_1 S + \sum_{i=2}^N \left( p_i \sum_{k=i}^N
     \frac{1}{\sigma(y_k)^2} \right),
\end{eqnarray}
and using a similar derivation,
\begin{eqnarray}
S_{xy} & = & \sum_{i=1}^N \frac{x_i y_i}{\sigma(y_i)^2} \\ 
   & = & y_1 \sum_{i=1}^N \frac{x_i}{\sigma(y_i)^2} + \sum_{i=2}^N
      \left( p_i \sum_{k=i}^N \frac{x_k}{\sigma(y_k)^2} \right) \\ 
   & = & y_1 S_x + \sum_{i=2}^N \left( p_i \sum_{k=i}^N
      \frac{x_k}{\sigma(y_k)^2} \right).
\end{eqnarray}
The standard equations (\ref{eq_a_random} \& \ref{eq_b_random}) can
then used to determine the best fit values of $a$ and $b$ for the
case of correlated uncertainties.  In fact, the values of $a$ and $b$
derived assuming correlated or uncorrelated uncertainties are exactly
the same.  The differences between the two types of uncertainties
arises in determining $\sigma_a$ and $\sigma_b$.

To derive $\sigma_a$ and $\sigma_b$ for a data ramp with
correlated measurements we start with equations 6.19 and 6.20 of
\citet{bev92}.  Converting from $y_i$ to $p_i$ as the
independent variable gives
\begin{eqnarray}
\sigma(cor)_z^2 & = & \sum_{i=1}^N \left[ \sigma(y_i)^2 
     \left( \frac{\partial z}{\partial y_i} \right)^2 \right] \\ 
  & = & \sum_{i=2}^N \left[ \sigma(p_i)^2 \left(
     \frac{\partial z}{\partial p_i} \right)^2 \right] \\
\end{eqnarray}
where $z$ is either $a$ or $b$.  The partial derivatives needed are
then
\begin{eqnarray}
\frac{\partial a}{\partial p_i} & = & \frac{1}{\Delta}
     \left( S_{xx} \frac{\partial S_y}{\partial p_i} - S_x
     \frac{\partial S_{xy}}{\partial p_i} \right) \\ 
  & = & \frac{1}{\Delta} \left( S_{xx} \sum_{k=i}^N
     \frac{1}{\sigma(y_k)^2} - S_x \sum_{k=i}^N
     \frac{x_k}{\sigma(y_k)^2} \right) 
\end{eqnarray}
and
\begin{eqnarray}
\frac{\partial b}{\partial p_i} & = & \frac{1}{\Delta}
     \left( S \frac{\partial S_{xy}}{\partial p_i} - S_x \frac{\partial
     S_y}{\partial p_i} \right) \\ 
  & = & \frac{1}{\Delta} \left( S \sum_{k=i}^N
     \frac{x_k}{\sigma(y_k)^2} - S_x \sum_{k=i}^N 
     \frac{1}{\sigma(y_k)^2} \right).
\end{eqnarray}
Thus,
\begin{eqnarray}
\sigma_a(cor)^2 & = & \sum_{i=2}^N
       \frac{\sigma(p_i)^2}{\Delta^2} \left( S_{xx} \sum_{k=i}^N
       \frac{1}{\sigma(y_k)^2} - S_x \sum_{k=i}^N
       \frac{x_k}{\sigma(y_k)^2} \right)^2~~\mathrm{and} \\
\sigma_b(cor)^2 & = & \sum_{i=2}^N
        \frac{\sigma(p_i)^2}{\Delta^2} \left( S \sum_{k=i}^N
        \frac{x_k}{\sigma(y_k)^2} - S_x \sum_{k=i}^N
        \frac{1}{\sigma(y_k)^2} \right)^2.
\end{eqnarray}

Finally, the uncertainties of the linear fit parameters for fits to
data with both correlated and random uncertainties (non-destructively
readouts with read noise) are
\begin{eqnarray}
\sigma_a^2 & = & \sigma_a(ran)^2 + \sigma_a(cor)^2~~\mathrm{and}
   \label{eq_sigma_a} \\
\sigma_b^2 & = & \sigma_b(ran)^2 + \sigma_b(cor)^2 \label{eq_sigma_b} .
\end{eqnarray}

\begin{figure}[tbp]
\plottwo{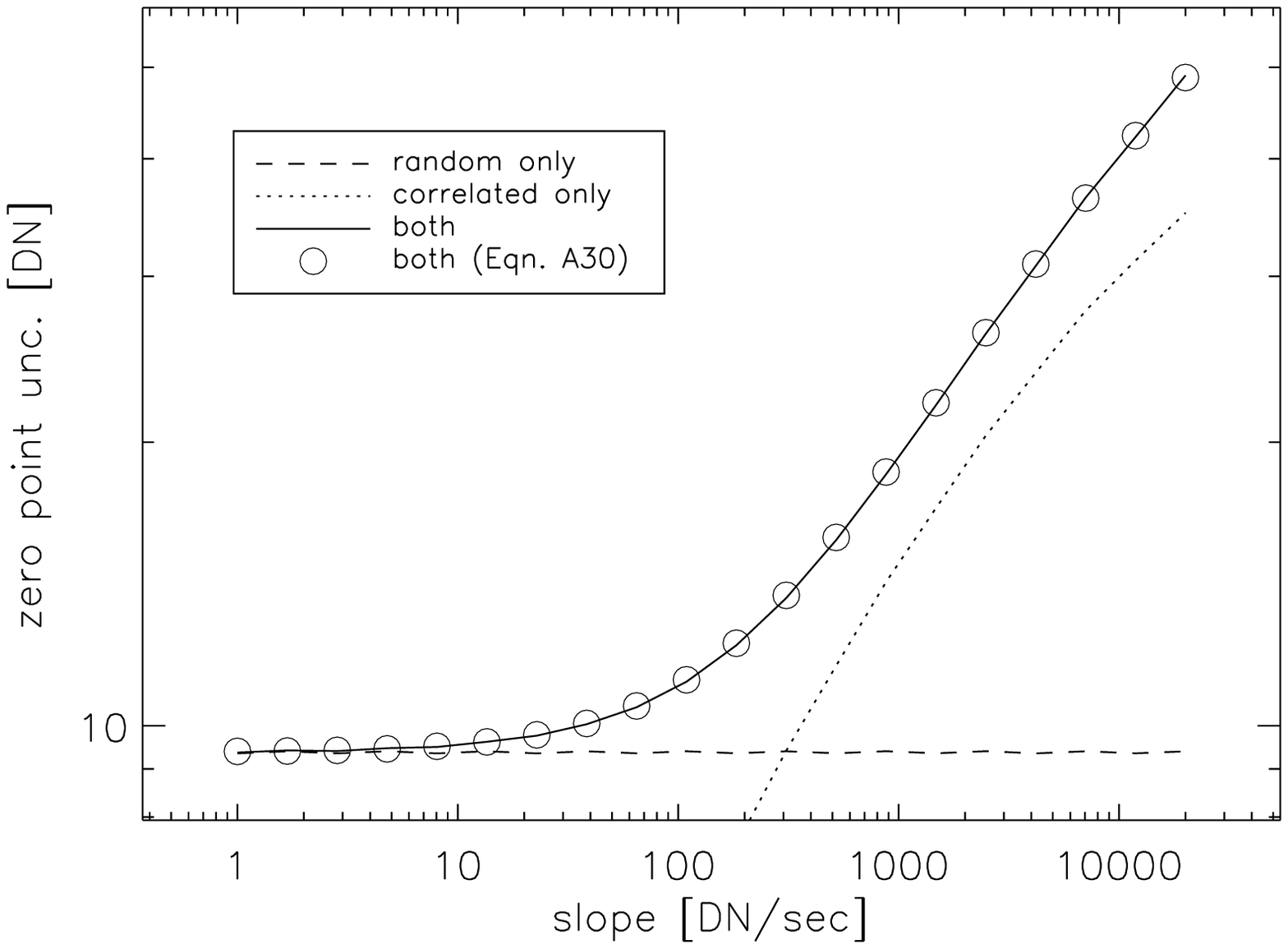}{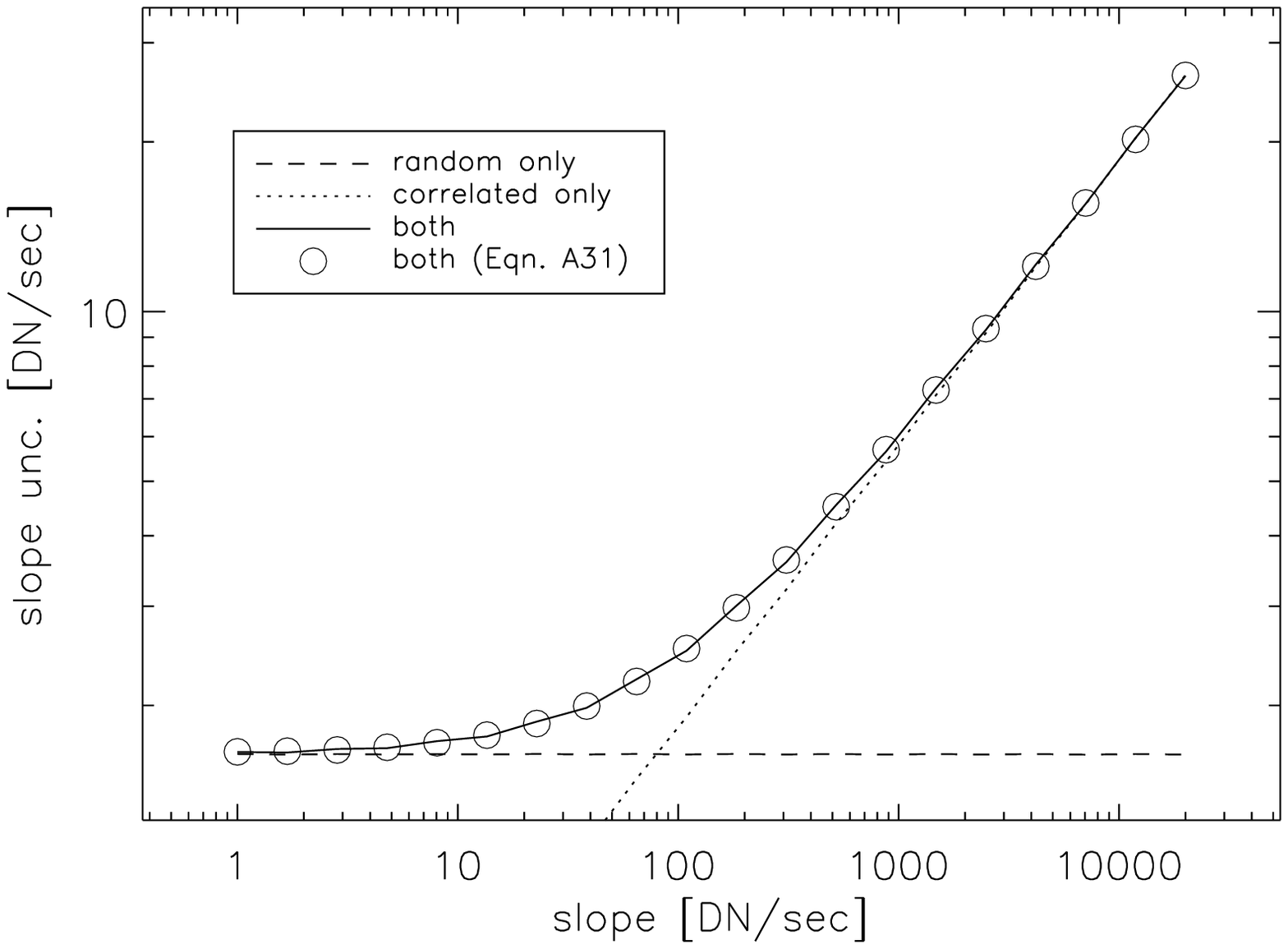}
\caption{The results of Monte Carlo simulations to test
Eqns.~\ref{eq_sigma_a} and \ref{eq_sigma_b} are plotted for the linear
fit zero point (left) and slope (right).  These particular Monte Carlo
runs were for cases similar to that expected for the 70~\micron\
array.  The line had a zero point of 3,000 DN, and exposure time of 10
seconds (80 reads) and a range of slopes (x axis).  Each data point
represents 10,000 trials.  The random only (dashed line), correlated
only (dotted line), and both (solid line) data give the true
uncertainties determined directly from the 10,000 trials.  The open
circles give the combined uncertainty using Eqns.~\ref{eq_sigma_a} and
\ref{eq_sigma_b}. \label{fig_fit_cor_ex}}
\end{figure}

The assumption that the uncertainties can be calculated separately
for the correlated and random measurement uncertainties was tested via
Monte Carlo simulations.  Simulations for cases similar to that
expected for the 70~\micron\ array are plotted in
Figure~\ref{fig_fit_cor_ex}.  As can be seen from these plots,
equations~\ref{eq_sigma_a} and \ref{eq_sigma_b} give very good
estimates of the actual uncertainties.


\begin{thebibliography}{}
\bibitem[Beeman \& Haller(2002)]{bee02} Beeman, J.~W.~\& 
   Haller, E.~E.\ 2002, \procspie, 4486, 209
\bibitem[Beichman et al.(1988)]{bei88} Beichman, C.~A., Neugebauer,
   G., Habing, H.~J., Clegg, P.~E., \& Chester, T.~J.\ 1988, NASA
   RP-1190, Vol.~1
\bibitem[Bevington \& Robinson(1992)]{bev92} Bevington, P.\ R.\ \&
   Robinson, D.\ K.\ 1992, Data Reduction and Error Analysis for the
   Physical Sciences (New York: McGraw-Hill, Inc.)
\bibitem[Burgdorf et al.(1998)]{bur98} Burgdorf, M.\ J., et al.\
   1998, Adv. In Space Research, 21, 5 
\bibitem[Church et al.(1993)]{chu93} Church, S.\ E., Griffin, M.\
   J., Price, M.\ C., Ade, P.\ A., Emergy, R.\ J., and Swinyard, B.\
   M.\ 1993, Proc.\ SPIE, 1946, 116
\bibitem[Cohen et al.(1992)]{cohen92} Cohen, M., Walker, R. G.,
  Barlow, M. J., \& Deacon, J. G. 1992, \aj, 104, 1650  
\bibitem[Cohen et al.(1995)]{cohen95} Cohen, M., Witteborn, F. C.,
  Walker, R. G., Bregman, J. D., \& Wooden, D. H. 1995, \aj, 110, 275
\bibitem[Cohen et al.(1996a)]{cohen96a} Cohen, M., Witteborn, F. C.,
  Carbon, D. F., Davies, J. K., Wooden, D. H., \& Bregman,
  J. D. 1996a, \aj, 112, 2274
\bibitem[Cohen et al.(1996b)]{cohen96b} Cohen, M., Witteborn, F. C.,
  Bregman, J. D., Wooden, D. H., Salama, A., \&Metcalfe, L. 1996b,
  \aj, 112, 241
\bibitem[de Graauw et al.(1996)]{deg96} de Graauw, T.~et al.\ 1996,
   \aap, 315, L49  
\bibitem[Dierckx et al.(1992)]{die92} Dierckx, B., Vermeiren, J.,
   Cos, S., Faymonville, R., and Lemke, D.\ 1992, Proc.\ ESA Symposium
   on Photon Detectors for Space Astronomy (SEE N94-15025), pp.\ 405 -
   408
\bibitem[Engelke(1992)]{engelke92} Engelke, C.~W.\ 1992, \aj, 104,
   1248  
\bibitem[Gordon et al.(2004)]{gor04} Gordon, K.~D.\ et al.\ 2004,
   \procspie, 5487, 177
\bibitem[Greisen \& Calabretta(2002)]{gre02} Greisen, E.~W.~\&
   Calabretta, M.~R.\ 2002, \aap, 395, 1061  
\bibitem[Haegel et al.(1999)]{hae99} Haegel, N.~M., Simoes, J.~C.,
   White, A.~M., \& Beeman, J.~W.\ 1999, \ao, 38, 1910 
\bibitem[Haegel et al.(2001)]{hae01} Haegel, N.~M., Schwartz, W.~R.,
   Zinter, J., White, A.~M., \& Beeman, J.~W.\ 2001, \ao, 40, 5748  
\bibitem[Heim et al.(1998)]{hei98} Heim, G.~B.~et al.\ 1998,
   \procspie, 3356, 985  
\bibitem[Heras et al.(2000)]{her00} Heras, A.~M., et al.\ 2000,
   Experimental Astronomy, 10, 177  
\bibitem[Hesselroth et al.(2000)]{hes00} Hesselroth, T., Ha, E.~C.,
   Pesenson, M., Kelly, D.~M., Rivlis, G., \& Engelbracht, C.~W.\
   2000, \procspie, 4131, 26 
\bibitem[Kessler et al.(1996)]{kes96} Kessler, M.~F.~et al.\ 1996,
   \aap, 315, L27 
\bibitem[Krist(1993)]{kri93} Krist, J.\ 1993, ASP Conf.~Ser.~52:
   Astronomical Data Analysis Software and Systems II, 2, 536  
\bibitem[Lemke et al.(1996)]{lem96} Lemke, D., et al. 1996, \aap,
   315, L64 
\bibitem[Low et al.(1984)]{low84} Low, F.~J., Beichman, C.~A.,
   Gillett, F.~C., Houck, J.~R., Neugebauer, G., Langford, D.~E.,
   Walker, R.~G., \& White, R.~H.\ 1984, Optical Engineering, 23, 122
\bibitem[Neugebauer et al.(1984)]{neu84} Neugebauer, G., et al.\ 1984,
   \apjl, 278, 1 
\bibitem[Rieke(2002)]{rie02} Rieke, G.~H. 2002, Detection of Light,
   2nd edition (Cambridge, England: Cambridge University Press)
\bibitem[Rieke et al.(1981)]{rie81} Rieke, G.~H., Montgomery, E.~F.,
   Lebofsky, M.~J., \& Eisenhardt, P.~R.\ 1981, \ao, 20, 814 
\bibitem[Rieke, Lebofsky, \& Low(1985)]{rieke85} Rieke, G. H.,
   Lebofsky, M. J., \& Low, F. J. 1995, \aj, 90, 900
\bibitem[Rieke et al.(2004)]{rie04} Rieke, G.~H., et
   al.\ 2004, \apjs, 154, 25
\bibitem[Schnurr et al.(1998)]{sch98} Schnurr, R., Thompson, C.~L.,
   Davis, J.~T., Beeman, J.~W., Cadien, J., Young, E.~T., Haller,
   E.~E., \& Rieke, G.~H.\ 1998, \procspie, 3354, 322  
\bibitem[Sparks(1998)]{spa98} Sparks, W.\ B.\ 1998, NICMOS
  Instrument Science Report, Space Telescope Science Institute, 98-008  
\bibitem[Swinyard et al.(1996)]{swi96} Swinyard, B.\ M.\ et al.\
   1996, \aap, 315, 43 
\bibitem[Swinyard et al.(2000)]{swi00} Swinyard, B., Clegg, P., Leeks,
   S., Griffin, M., Lim, T., \& Burgdorf, M.\ 2000, Experimental
   Astronomy, 10, 157  
\bibitem[Valentijn \& Thi(2000)]{val00} Valentijn, E.~A.~\& Thi,
   W.~F.\ 2000, Experimental Astronomy, 10, 215  
\bibitem[Young et al.(1998)]{you98} Young, E.~T.~et al.\ 1998,
   \procspie, 3354, 57  
\bibitem[Young et al.(2003)]{you03} Young, E.~T.~et al.\ 2003,
   \procspie, 4850, 98  
\end{thebibliography}
\end{document}